\documentclass[manuscript]{acmart}

\pdfoutput=1
\usepackage{dirtytalk}
\usepackage{graphicx}
\usepackage{subfigure}
\usepackage{multirow}
\usepackage{calc}
\usepackage{float}

\usepackage{textcomp}
\usepackage{siunitx}
\usepackage{soul,xcolor}

\AtBeginDocument{%
  \providecommand\BibTeX{{%
    \normalfont B\kern-0.5em{\scshape i\kern-0.25em b}\kern-0.8em\TeX}}}

\definecolor{burgundy}{RGB}{144,0,32}

\acmDOI{xx.xxxx/xxxxxxx.xxxxxxxx}

\begin{document}

\setstcolor{red}



\title[ShareYourReality]{ShareYourReality: Investigating Haptic Feedback and Agency in Virtual Avatar Co-embodiment}


\author{Karthikeya Puttur Venkatraj}
\affiliation{
  \institution{Centrum Wiskunde \& Informatica}
  \institution{Delft University Technology}
  \city{Amsterdam}
  \country{Netherlands}}
\email{karthike@cwi.nl}
\orcid{0009-0003-4245-8802}

\author{Wo Meijer}
\affiliation{
  \institution{Delft University of Technology}
  \city{Delft}
  \country{Netherlands}}
\email{W.I.M.T.Meijer@tudelft.nl}
\orcid{0000-0002-8369-6394}

\author{Monica Perusquía-Hernández}
\affiliation{
  \institution{Nara Institute of Science and Technolgy}
  \city{Ikoma}
  \country{Japan}}
\email{m.perusquia@is.naist.jp}
\orcid{0000-0002-0486-1743}

\author{Gijs Huisman}
\affiliation{
  \institution{Delft University of Technology}
  \city{Delft}
  \country{Netherlands}}
\email{g.huisman@tudelft.nl}
\orcid{0000-0002-8029-5042}

\author{Abdallah El Ali}
\affiliation{
  \institution{Centrum Wiskunde \& Informatica}
  \city{Amsterdam}
  \country{Netherlands}}
\email{aea@cwi.nl}
\orcid{0000-0002-9954-4088}

\renewcommand{\shortauthors}{Venkatraj et al.}

\begin{abstract}

Virtual co-embodiment enables two users to share a single avatar in Virtual Reality (VR). During such experiences, the illusion of shared motion control can break during joint-action activities, highlighting the need for position-aware feedback mechanisms. Drawing on the perceptual crossing paradigm, we explore how haptics can enable non-verbal coordination between co-embodied participants. In a within-subjects study (20 participant pairs), we examined the effects of vibrotactile haptic feedback (None, Present) and avatar control distribution (25-75\%, 50-50\%, 75-25\%) across two VR reaching tasks (Targeted, Free-choice) on participants’ Sense of Agency (SoA), co-presence, body ownership, and motion synchrony. We found (a) lower SoA in the free-choice with haptics than without, (b) higher SoA during the shared targeted task, (c) co-presence and body ownership were significantly higher in the free-choice task, (d) players’ hand motions synchronized more in the targeted task. We provide cautionary considerations when including haptic feedback mechanisms for avatar co-embodiment experiences. 

\end{abstract}

\begin{teaserfigure}
        \centering
        \setlength{\abovecaptionskip}{2pt}
        \includegraphics[width=1\columnwidth]{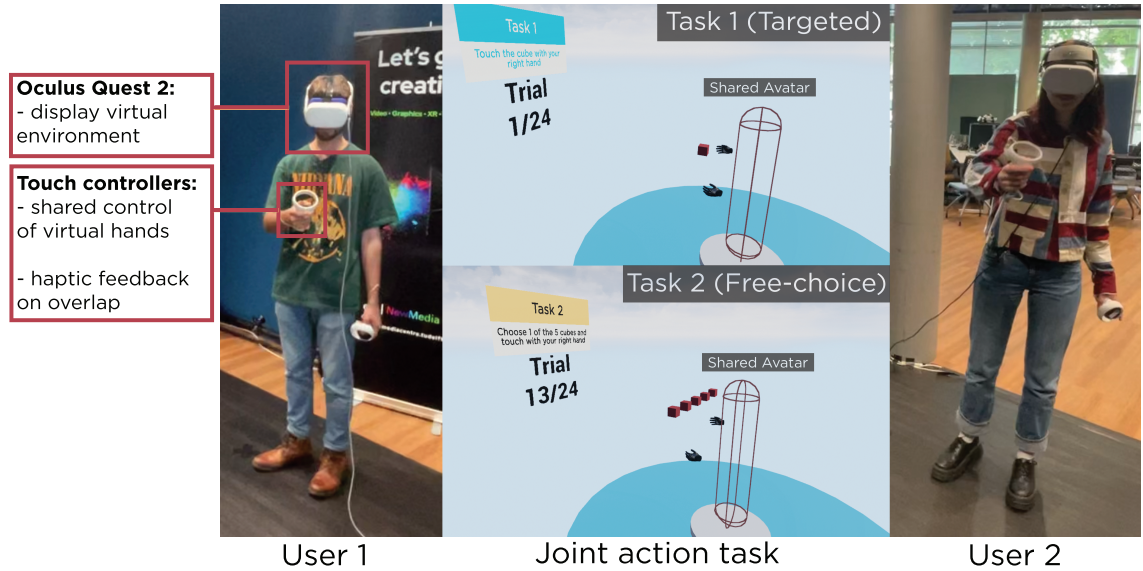}
                \caption{Two users co-embodying virtual hands performing two types of joint action tasks in a shared virtual reality environment}                
        \label{fig:teaser}
        \Description{Four images showing the setup of the experiment. On the left an image of User 1 wearing oculus quest 2 head mounted displays and holding touch controllers while performing reaching motion and on the right an image of user 2 with the same setup; In the center (top) an image of the setup of Task 1 (targeted) with a single cube and shared avatar and (bottom) an image of the setup of task 2 (Free-choice) with five cubes and shared avatar} 
    \end{teaserfigure}



    

\maketitle

\section{Introduction}

Virtual Reality (VR) technologies enable people to not only immerse themselves in artificial digital worlds, but also enable previously impossible interactions that challenge our assumptions about our virtual bodies and social coordination processes in such virtual spaces. The proliferation of consumer head-mounted displays (HMDs) have made VR systems an increasingly common platform through which virtual social interactions can take place~\cite{Hamad2022,Tekla2016}. VR enables meeting in a shared, immersive virtual environment~\cite{heidicker_influence_2017,Li2019a}, and interacting with virtual representations of human avatars and virtual agents. Social interactions are a key factor in VR -- not only to prevent isolation of individuals in the virtual environment~\cite{rizvic_da_2022}, but also to enable joint social activities and interactions~\cite{rasch_going_2023, theodoropoulos_developing_2023} in and beyond social VR platforms such as VRChat and Rec Room~\cite{oh_systematic_2018}. 

Despite the myriad ways that social VR platforms can implement multi-user functionality, the most commonly used method today is giving each user their own individual avatar to navigate through the virtual environment. Typically, such interactions adopt a first-person perspective, which is an important contributing factor to creating a sense of body ownership (i.e., the experience of having a virtual body~\cite{kilteni_sense_2012}) and agency (i.e., involving a sense of control over a virtual body~\cite{Bennett2023agency}) towards the user's virtual avatar. Extending beyond common social VR interactions, researchers have recently explored the concept of “virtual co-embodiment”~\cite{hagiwara_shared_2019, fribourg_virtual_2021}, where two users embody a single, shared avatar. This inherently differs from a shared visual experience where multiple users would only share the same viewing perspective~\cite{yang_2002}. Virtual co-embodiment offers a multi-user experience characterized by shared control over the avatar's movement. Since two or more people share control over the avatar, there is an increase in social coordination~\cite{fribourg_virtual_2021}. Such `fusionary' interactions are a component of what is dubbed the ``JIZAI Body"~\cite{Inami2022}: the concept by which a computer-mediated human body can seamlessly adapt to social structure changes, such that any additions or alterations (virtual or physical) would feel as much their own as their original body. This relates to recent efforts toward sensible human-computer integration~\cite{Mueller2020,Cornelio2022}, which extends the notion of cybernetics~\cite{Wiener2008}. Indeed, experiments have shown that participants who co-embody a virtual avatar report high levels of perceived control, with lower levels of actual control~\cite{hapuarachchi_effect_2023, kodama_enhancing_2022}. This enhanced perception of control can be useful for rehabilitation specialists and support personnel that use immersive technology for the treatment of physical and cognitive function of individuals that have suffered from stroke~\cite{mekbib2021novel} and dementia~\cite{zhu2021dementia}, respectively. The shared control of virtual co-embodiment provides an assistive methodology to improve accessibility for these vulnerable individuals during the setup and navigation of virtual environments. In cases of stroke rehabilitation, co-embodiment can also be incorporated alongside techniques such as Mirror Therapy~\cite{lin2021development}, to enrich motor support in the VR system by incorporating shared (human) assistance to adapt to requirements for each individual need.

Research has shown that individuals can adapt to different media to achieve their communication goals~\cite{papacharissi_real-virtual_2005}, where the degree of embodiment and time required for such embodiment may vary depending on the visual and visuo-motor consistency of the artificial (virtual) and biological body \cite{Argelaguet2016,fribourg_avatar_2020,Keenaghan2020body}. Furthermore, when another entity that is sentient or appears to be sentient is present in the same environment, another dimension called ‘social presence’ comes into play~\cite{oh_systematic_2018,Harrison2018}. The degree of experienced social presence depends on the person's perceived ability to access another individual's intentions, intelligence, and sensory impressions. Immersive qualities, contextual properties, and individual differences, can predict the extent to which social presence is experienced by users in VR~\cite{oh_systematic_2018}. Without a sufficient level of social presence, the other entity is perceived as artificial and not as an intentional social being~\cite{oh_systematic_2018}. Without such perceived intentionality, shared and collaborative tasks become difficult, making a high sense of social presence vital for a smooth collaborative experience to occur during virtual co-embodiment.

However, a specific challenge in co-embodiment is that the visual feedback alone of the combined motion of the shared avatar does not fully provide the intentional information of one's partner because the shared avatar's motion is partially determined by one's own motion. Virtual co-embodiment leads to a situation where there is an intermingling of self-presence with social presence, where the identity of the self is intrinsically linked to the avatar and the presence of another intentional subject. This intentionality is translated through the amount of control available for each person over the shared avatar. While researchers have explored virtual co-embodiment in relation to users’ perception of their embodiment of the shared avatar~\cite{hagiwara_shared_2019, fribourg_virtual_2021} and its use for motor skill learning~\cite{kodama_enhancing_2022, kodama_effects_2023}, the role of social presence and its influence on social coordination within this context has not yet been fully characterized. Extensions of this concept by~\citet{hapuarachchi_knowing_2022} and \citet{hapuarachchi_effect_2023} have highlighted the need for non-verbal communication mechanisms to be implemented in the paradigm of co-embodiment to enhance sense of embodiment towards the co-embodied avatar. Indeed, during such co-embodiment experiences, the illusion of shared motion control can be coupled with user communication techniques that support coordination
, highlighting the need for position-aware feedback mechanisms~\cite{vesper2017joint}. Haptic feedback may be especially appropriate here because it can positively influence the experience of social presence~\cite{oh_systematic_2018}. For this, we draw on \citet{auvray_perceptual_2009}'s haptic 'perceptual crossing paradigm', which was conceived to study social interaction dynamics in real time through tactile sensorimotor interactions~\cite{kojima2017sensorimotor}. Perceptual crossing refers to situations where two perceptual activities of the same kind meet each other, such as when two people catch each other’s eye (joint gaze and attention \cite{Stephenson2021}), or mutual social touch \cite{SAARINEN2021}. This paradigm offers us a foundation to build systems that enable studying of factors involved in mutual recognition between people in remote interactions. Given the focus of co-embodiment scenarios on a sense of shared control with a remotely located other person, we believe the perceptual crossing paradigm lends itself well to studying sensorimotor interactions in co-embodied VR.


In this paper, we draw on the perceptual crossing paradigm \cite{auvray_perceptual_2009}, to explore how haptic feedback can be integrated into a virtual co-embodiment scenario, where pairs of participants share control over a virtual hand. Building on the idea of haptic perceptual crossing, we implemented haptic feedback (on/off) when users' hand positions overlap in virtual space. We examine how haptic cues can enable awareness of each other and coordination between co-embodied participants. We ask: \textbf{(RQ)} How do haptic feedback mechanisms and varied avatar control distribution influence users' sense of agency, co-presence, body ownership, and motion synchrony in targeted and free-choice virtual avatar co-embodiment tasks? In a controlled, within-subjects study with 20 participant pairs, we examined the effects of positional haptic feedback (None, Present) and avatar control distribution (25-75\%, 50-50\%, 75-25\%) across two cube selection tasks (Targeted, Free-choice) on participants' sense of agency (SoA), co-presence, body ownership, and motion synchrony. Our findings showed (a) a lower sense of agency in the free-choice with haptics compared to no haptic feedback, (b) higher agency during the shared target task, (c) co-presence and embodiment were significantly higher in tasks where there were multiple targets, and (d) players' hand motions synchronized more in the targeted task. 

Our exploratory work offers two primary contributions: \textbf{(1)} We integrate the concept of perceptual crossing into the paradigm of virtual co-embodiment to enable position-aware haptic non-verbal communication cues between two users; \textbf{(2)} We provide empirically backed insights showing the influence of haptics on the perceived sense of agency during targeted and free-choice selection tasks under variable control of shared virtual hands.

\begin{figure}[t]
    \centering
    \includegraphics[width=0.7\columnwidth]{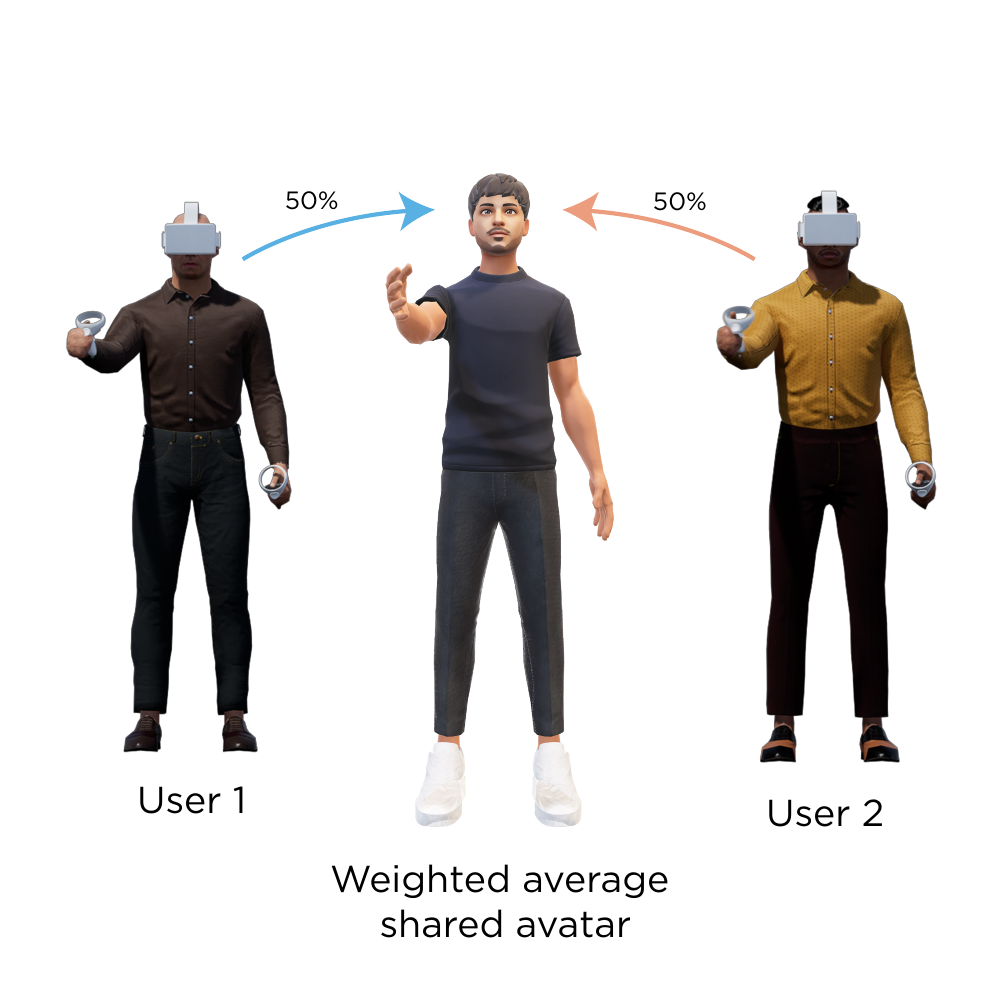}
    \caption{Illustration of the weighted average virtual co-embodiment method, where the motion of the shared avatar (center) is generated by taking the weighted average of the motion of User 1 (left) and User 2 (right)}
    \Description{An illustration of User 1 (left) and User 2 (right) embodying a virtual shared avatar with its motion generated by taking the weighted average of both users' motions}
    \label{fig:coembodiment}
\end{figure}

\section{Background and related Work}
In this section, we describe prior work on virtual co-embodiment, control sharing techniques, haptics integrated into co-embodiment experiences, and the perceptual crossing paradigm.

\subsection{Virtual avatar co-embodiment and the Sense of Agency}
\label{sec:rel_work}

Virtual co-embodiment refers to occurrences where multiple users can simultaneously interact with the virtual environment using a shared avatar. Given that two or more individuals share control over the avatar, there is an increase in social coordination~\cite{fribourg_virtual_2021}. To this end, prior work has shown that participants who co-embody a virtual avatar reported high levels of perceived control, with lower levels of weighted percentage of actual control~\cite{hapuarachchi_effect_2023, kodama_enhancing_2022}. This makes it a promising tool for VR-based rehabilitation~\cite{juan2023immersive} and training~\cite{gonzalez-franco_immersive_2017, harris_exploring_2023} applications, since a learner with low control can feel a stronger sense of agency (SoA) while performing the activity with a teacher with high control. One domain in which researchers are trying to leverage the immersive capabilities of VR is in the support and treatment of dementia patients~\cite{zhu2021dementia} -- these individuals are considered vulnerable and they typically find it challenging to operate basic VR controls~\cite{amanda2014rt}. In such cases, co-embodiment can enable assistive accessibility for these individuals, guided by support personnel. This would enable them to not only train, but also maintain a high level of agency during such immersive experiences. Furthermore, VR has shown potential for its ability to stimulate Mirror Neurons (MNs) of the internal sensorimotor system of stroke patients~\cite{carvalho2013mirror}. In such settings, patients are immersed in training scenarios in virtual environments that involve executing motor actions, such as observing and visualizing mirror limb movements with the intent to imitate these actions. These have shown enhanced MN activation, leading to faster post-stroke recovery~\cite{mekbib2021novel}. To that end, co-embodiment can be further leveraged within these techniques in order to improve the effectiveness of such treatments through enhanced agency and nuanced control over movements executed by the patients actively guided by their trainers or caregivers.

In the context of avatar co-embodiment, the ‘Sense Of Embodiment’ (SoE) can be manifested through three main components: Sense of Self-Location, Sense of Body Ownership, and the Sense of Agency~\cite{kilteni_sense_2012}. Sense of Self-Location refers to the feeling of ‘being inside’ a virtual body, while sense of body ownership and agency refers to the feeling of ‘having’ and ‘controlling’ the virtual body, respectively. Studies have explored various factors and their influence on these components, and have shown that manipulations of the overall SoE are possible through changes in avatar representations, degree of control, and perspective of the users~\cite{fribourg_avatar_2020, ogawa_virtual_2019}. Similarly, the influence of sharing the virtual body with another user and its effect on SoE was studied in experiments of virtual co-embodiment. Here, the sense of agency and body ownership play a pivotal role that determines the engagement level during the shared perceptual activity~\cite{hagiwara_shared_2019, fribourg_virtual_2021, hapuarachchi_knowing_2022, hapuarachchi_effect_2023, kodama_effects_2023}.

\subsection{Avatar co-embodiment control sharing techniques}

To realize such co-embodiment, the motion of a shared avatar has previously been generated using two techniques: the weighted average co-embodiment method~\cite{hagiwara_shared_2019,fribourg_virtual_2021, kodama_enhancing_2022} and the body-part-segmented co-embodiment method~\cite{hapuarachchi_knowing_2022,hapuarachchi_effect_2023}. The weighted-average method involves assigning a weight between 0 and 100 percent to each user and generating the movement of the shared avatar by interpolating the weighted average of the real-time position and orientation of the controllers of both the users (Figure~\ref{fig:coembodiment})~\cite{hagiwara_shared_2019, fribourg_virtual_2021, kodama_enhancing_2022}. The body-part-segmented co-embodiment method is a technique where the motion of independent limbs of a shared avatar is controlled by each user separately~\cite{hapuarachchi_knowing_2022, hapuarachchi_effect_2023}. In this paper we focus on the first method, where shared interactions can be manipulated by both the users and their influence is determined by the percentage of control they possess. Since we focus on position-aware feedback mechanisms, we draw on the weighted-average method to enable this. Results from~\cite{hagiwara_shared_2019, fribourg_virtual_2021, kodama_enhancing_2022} all showed that sense of agency increased with the increase in the control weight for the participant during avatar co-embodiment. In all the studies, participants could coordinate their movements in joint action, leading to the sharing of motor intention and synchronization. In a follow-up study by~\citet{kodama_effects_2023}, they evaluated participants' task performance and motor skill learning ability. They concluded that learning using virtual co-embodiment was more efficient than the perspective-sharing method, in which a translucent teacher avatar was superimposed on the learner's first-person perspective view. However, contrary to the previous studies, no significant differences were observed between the different control weight conditions. We draw on the design considerations from~\cite{hagiwara_shared_2019, fribourg_virtual_2021, kodama_enhancing_2022} to design our study protocol.

\subsection{Haptics for virtual co-embodied experiences}

Since the early days of VR, haptic feedback has been a central component in many VR systems~\cite{srinivasan1997haptics} and has been used to enable a diversity of touch-based interactions in VR ~\cite{wee_haptic_2021}, with the most common type of haptic feedback in VR applications being vibrotactile and force feedback~\cite{Yang_2021}. Studies that have explored haptics as a communication medium in the context of shared virtual spaces report enhanced user experiences~\cite{jung_use_2021, zhang_remotetouch_2023}. Important to our present purposes, the addition of haptic feedback to social VR has been found to consistently enhance perceived social presence \cite{oh_systematic_2018}. In work on co-embodiment, the application of haptic feedback is essentially understudied. \citet{hapuarachchi_knowing_2022} explored the manipulation of the sense of agency by providing visual feedback of the partner's target during co-embodiment, and \citet{hapuarachchi_effect_2023} implemented passive haptics by attaching a back brace to both the users, allowing them to maintain consistent shoulder posture while controlling the shared avatar using the body-part-segmented co-embodiment method. These explorations highlight the value of identifying what type of feedback modalities can be integrated into the virtual co-embodiment paradigm to provide users with advanced perceptual capabilities. While visual feedback offers more information to the user, it leads to cluttered, chaotic experiences when scaled up. Thus, haptic feedback provides an alternative to overcome this limitation. The challenge in the context of co-embodiment is to design the feedback mechanism in a way that does not increase the cognitive load required to differentiate between the interaction with the environment and the presence of the other user. Given the foregoing, we implement haptic feedback as a communication medium to indicate the other users' position during co-embodiment.


\subsection{Perceptual crossing paradigm for enhancing social coordination}
The perceptual crossing paradigm~\cite{auvray_perceptual_2009} was conceived to study social interaction dynamics in real time through tactile sensorimotor interactions~\cite{kojima2017sensorimotor}. The classical paradigm features a minimalist 1D environment--a line--that loops around creating a continuous interaction space not visible to the users. Two users are each represented in the virtual space by an avatar (a dot) that they can control using a standard computer mouse. When their avatar encounters a virtual object in the 1D space they receive haptic feedback. There are three types of objects in the environment: a static object, the user's avatars, and the user's 'shadow,' an object that moves with the users' avatars at a set distance. All objects have exactly the same size and produce the same haptic feedback when encountered. When one user encounters the shadow of the other user, only the interacting user receives haptic feedback while the other user (to whom the shadow belongs), does not. The only condition when both users receive haptic feedback simultaneously is when both users' avatars encounter each other. Users are tasked with clicking the mouse when they think they are interacting with the avatar of the other user. In the original studies~\cite{auvray_perceptual_2009,auvray2012perceptual}, users were successfully able to locate each other in the 1D virtual space. However, the probability of clicking when encountering another user's avatar was not significantly higher than clicking when encountering another user's shadow. Successful identification of the other could only be explained by the stability of mutual recognition; users would encounter one another, move back, encounter each other again, and repeat, creating an oscillating movement pattern of repeated encounters. In other words, users would only successfully recognize each other during perceptual crossing (i.e., perceptual activities of the same kind meet each other). 

Extensions of the basic paradigm have shown that, in a team-based version of the paradigm where users were instructed to collaborate, participants successfully identified the other's avatar and, for those encounters, reported the clearest awareness of the others' presence \cite{froese2014embodied}. A version of the paradigm that used a following task in a skewed 1D environment highlighted that users were successful in following each other's movements through haptic perceptual crossing \cite{lenay2021perceiving}. Though part of the strength of the perceptual crossing paradigm lies in the minimalist approach, 2D extensions have already been successful~\cite{lenay_you_2011}. To the best of our knowledge, no 3D implementations of the paradigm have ever been attempted. We see an interesting opportunity in the implementation of the paradigm's basic premise (i.e., haptic feedback upon contact in a virtual space to signify the presence of the other) as an interactive cue that could aid movement coordination as well as enhance perceived social presence in virtual co-embodiment scenarios.

\begin{figure*}[t]
    \centering
    \includegraphics[width=0.9\linewidth]{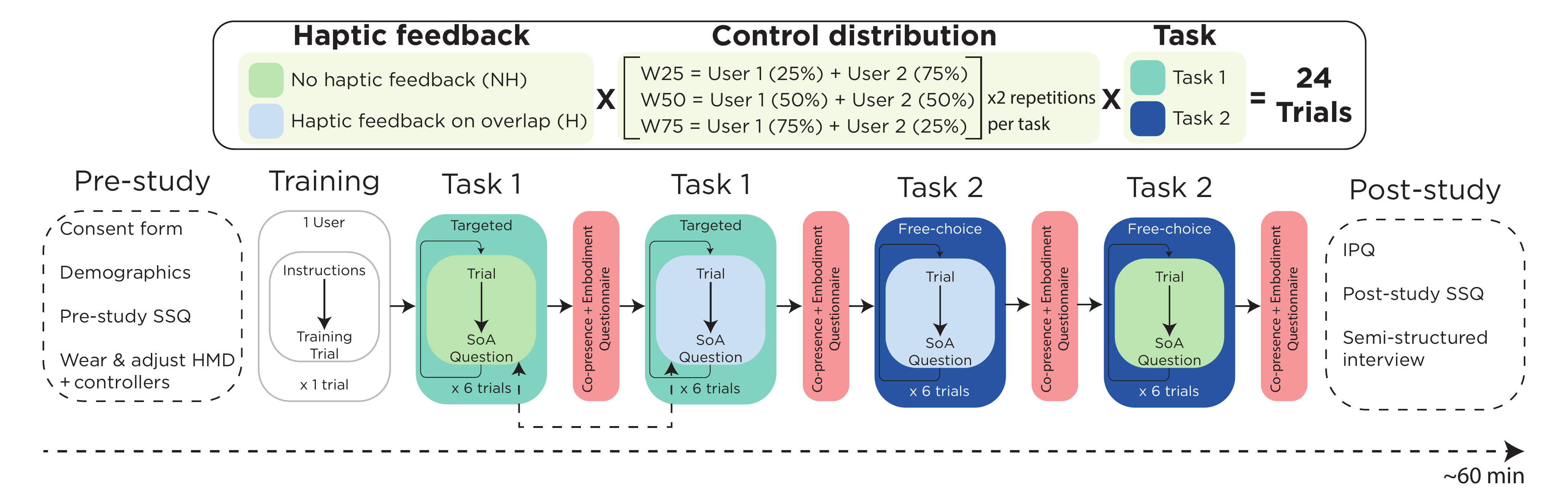}
    \caption{Diagram illustrating the different phases of the Study procedure, along with textual labels explaining each component}
    \label{fig:studydesign}
    \Description{Diagram illustrating the different phases of the Study procedure, along with textual labels explaining each component}
\end{figure*}

\section{Methods}
In this section, we describe our research methodology, including the study design, experimental protocol, objective and subjective measures, our hardware and software setup, study procedure, and participant sample.

\subsection{Study design}

Our study has two main Independent Variables (IVs), and follows a 3 (IV1: Control Distribution: 25-75\% vs. 50-50\% vs. 75-25\%) x 2 (IV2: Haptic Feedback: None vs. Present) within-subject design, tested in a controlled, virtual  environment. The control distribution consisted of three sets of weighing player one and player two's control over the shared avatar: 25-75\%, 50-50\%, 75-25\% (referred to as W25, W50, W75). There was either no haptic feedback (NH) or haptic feedback when participants' hands overlapped (H) in the virtual space. This interaction was designed using virtual spheres (radius 8 cm), which is approximately the size of the virtual hand mesh that was attached to the controllers in the virtual environment. When the spheres of each user intersect with each other, the haptic feedback is triggered for both the users. The study was divided into three phases: Training, Task 1, and Task 2 (Figure \ref{fig:studydesign}). There were six distinct conditions (3 control distribution x 2 haptic feedback) for each task that was performed by a pair of participants. The experiment consisted of two tasks where each of these conditions were repeated twice, bringing the total to 24 (2 tasks x 2 repetitions) trials for the entire study. The subjective responses of the participants on the sense of agency were collected with a questionnaire after each trial, while the sense of co-presence and embodiment questionnaires were collected after each haptic feedback condition / block (after six trials). Task 1 was always performed before Task 2, and the two haptic feedback conditions were counterbalanced according to a Latin square design such that starting trial of each session consisted of all possible combinations of haptic and control conditions, with the remainder of trials subsequently randomized to mitigate order effects. The study was designed such that a sample consists of pairs of participants. For example, in a session, participant 1 would perform the task with 25\% control four times (4x), twice with and twice without haptic feedback, while their counterpart had 75\% control (performed also 4x). Similarly for 50\% (4x) and 75\% (4x). Therefore for "25\%", "75\%", and "50\%", there were 4 samples for each task (twice with and twice without haptic feedback).

\subsection{Protocol}
\subsubsection{Joint action reaching tasks}
The most common method to evaluate virtual co-embodiment is using a reaching task. In this task, participants touch an object such as a cube~\cite{hagiwara_shared_2019} or sphere~\cite{fribourg_virtual_2021, hapuarachchi_knowing_2022} using a shared avatar. This task typically focuses only on participants' motion, as adding additional interactions (e.g., button presses) can increase task complexity, which may render the task unsuitable for studying shared control. \citet{fribourg_virtual_2021} introduced a reaching task, for three scenarios: free, target, and trajectory. During the free task, each participant was free to choose any sphere to touch, while the sphere to be touched was highlighted in the target task. The trajectory task involved following a particular path before touching a highlighted sphere, and it focused more on precision. To help answer our RQ, we need to better understand the influence of movement freedom and intention on the level of embodiment (sense of agency and body ownership) over the shared avatar using haptics. Therefore, we implemented two reaching tasks: targeted (Task 1) and free-choice (Task 2), which we describe below. 


 
\subsubsection{Training}
In the training phase, basic controls of our VR system were explained to each participant, including how to use the controller buttons to interact with widgets in the scene. Afterward, each participant performed an individual training trial, which showed the participants how to complete Task 1. Since the training session was performed individually, no haptic feedback was provided beforehand, since this can only occur in the later part of the study involving co-embodiment.  

\subsubsection{\textbf{Task 1}: Targeted}
\begin{figure}
	\centering
	\subfigure[Cube spawned in participant's field of view]{\label{fig:task1_1}\includegraphics[width=0.48\linewidth]{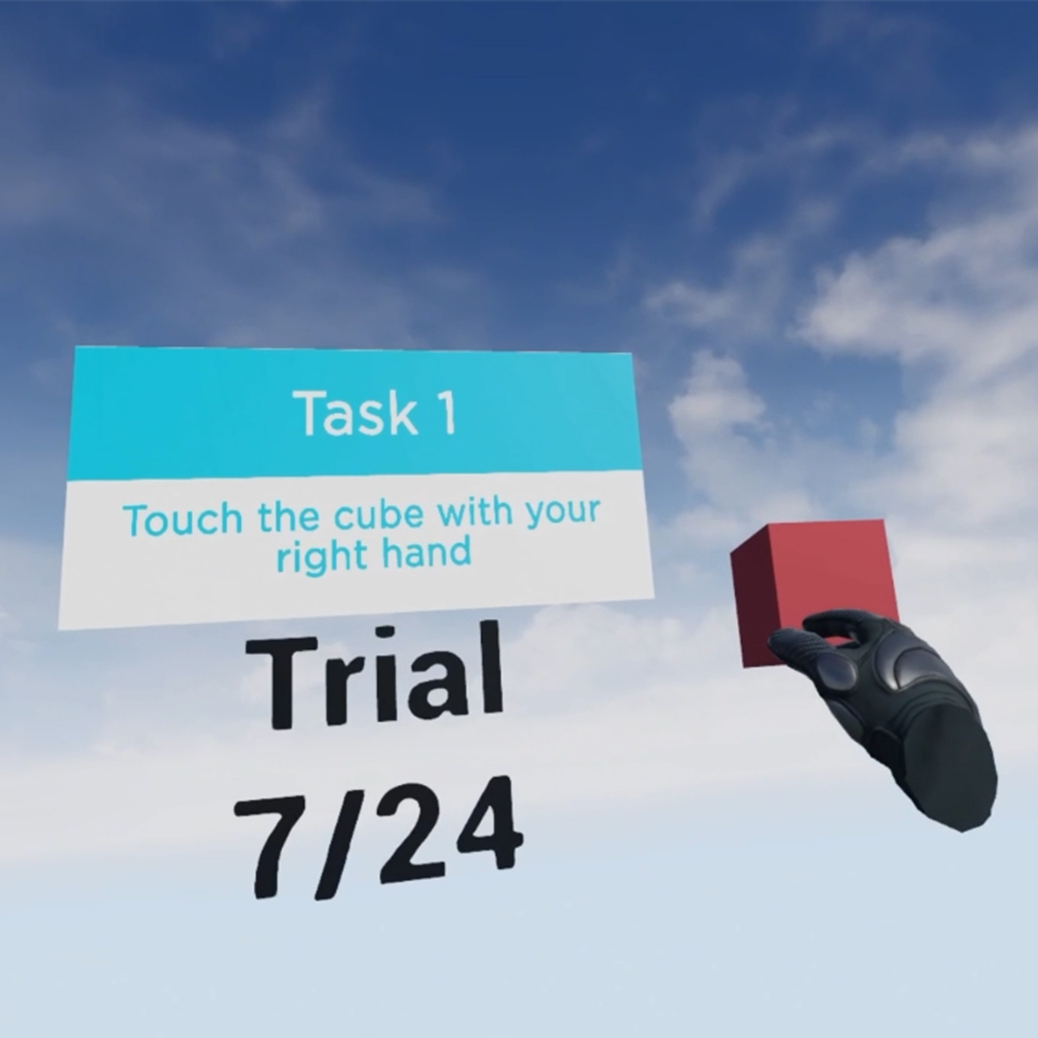}}
	\subfigure[Cube spawned in next position after collision with virtual hand]{\label{fig:task1_2}\includegraphics[width=0.48\linewidth]{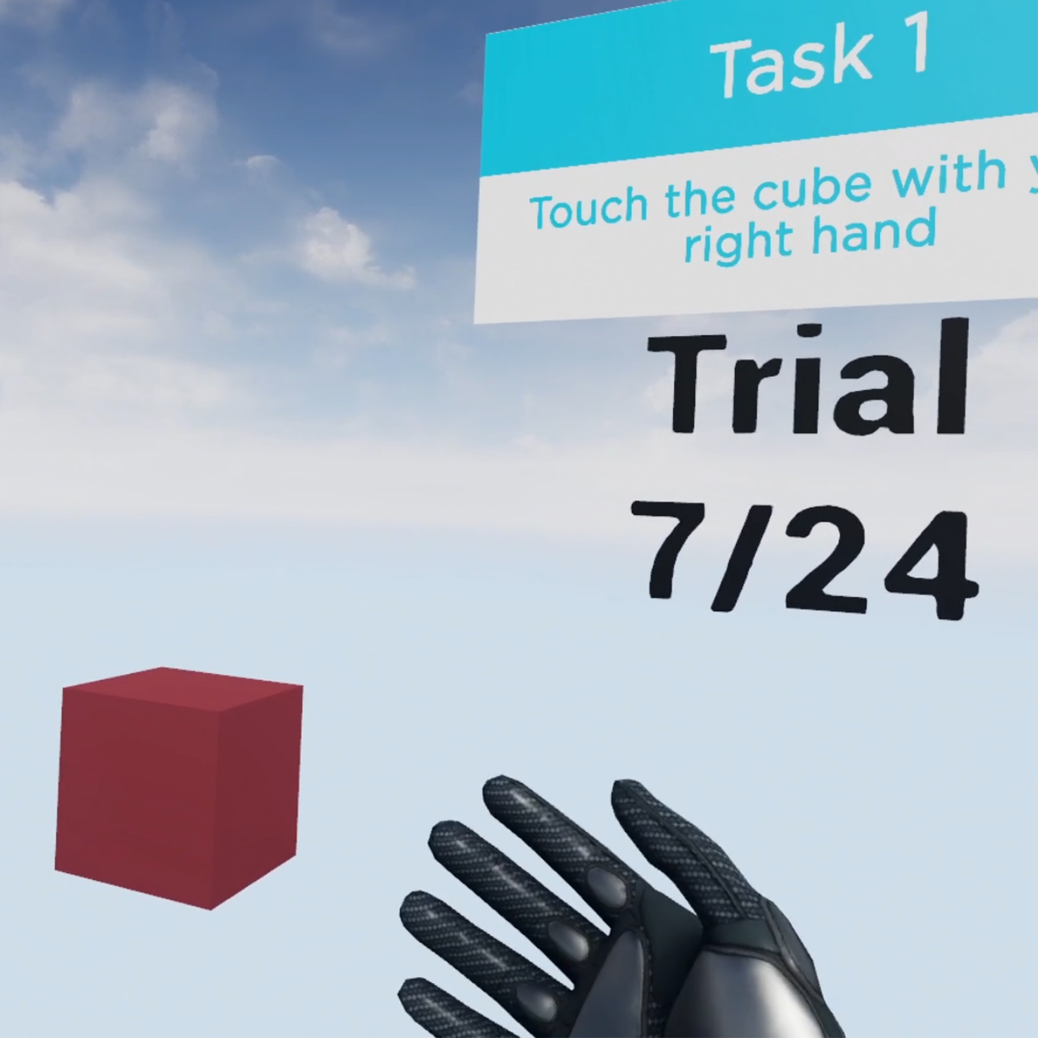}}
	\caption{First-person perspective of cube interaction in the targeted task}
	\Description{Two images of the first-person perspective of users performing Task 1 (targeted); on the left, the virtual shared hand is reaching out to touch the cube, and on the right, the cube re-spawns at a different location}
	\label{fig:task1}
\end{figure}

In Task 1, participants used the shared right hand of the avatar to touch a cube that spawned in their field of view (Figure \ref{fig:task1_1}). Once the shared hand collided with the cube, the cube would be removed. After a second delay, another cube was spawned at a pseudo-randomized location\footnote{The random locations were limited to within the space in front of the participants to ensure the cubes were visible and reachable.} (Figure \ref{fig:task1_2}). The location was pseudo-randomized, instead of pre-generated, to minimize learning effects. The delay provided a small reset time for the participants to avoid physical and cognitive fatigue. A spatial chime sound also originated from the spawned location of the cube, to indicate to participants the location of the new cube as it is difficult for users to realize if the cube has appeared at a location that is not within their field of view. Participants had to touch the cube a total of 17 times in each trial during Task 1.

\subsubsection{\textbf{Task 2}: Free-choice}
\begin{figure}
    \centering
    \includegraphics[width=0.9\linewidth]{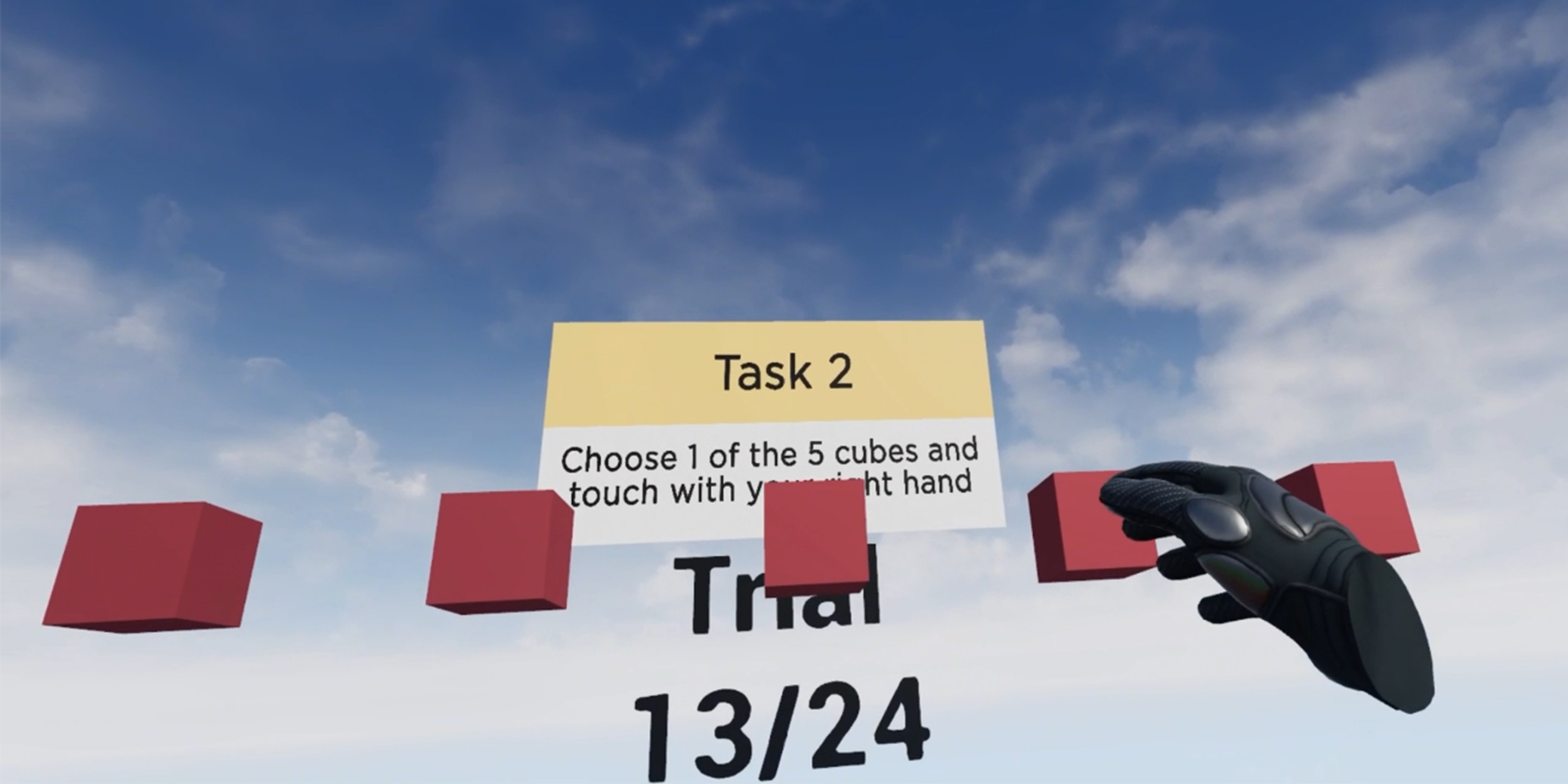}
    \caption{First-person perspective of cube interaction in the free-choice task}
    \label{fig:task2}
    \Description{First person perspective of users performing task 2 (free choice), where virtual shared hand is reaching out to touch one out of the five cubes in the view}
\end{figure}

In Task 2, five cubes are spawned in front of the participants, who had to move the shared hand to touch any of them to progress (Figure \ref{fig:task2}). In this case, when the shared hand collided with any of the cubes, all the cubes would get removed, and after a second delay, all the cubes would re-spawn back in the same positions. During Task 2, in each trial, participants had to touch one of the five cubes a total of five times to proceed. This task was designed to simulate a scenario where participants would have to collaboratively choose the co-embodied movement without verbal communication. This provided a suitable scenario to investigate if the position-aware feedback mechanism modeled on the perceptual crossing paradigm will enable the co-embodied users to work together.

\subsection{Measures}
\subsubsection{Objective measures}
The orientation (Roll, Pitch, and Yaw) and position (X, Y, Z coordinates) of both participants' HMD and controllers were recorded at the applications' default sampling rate of 70Hz during the entire session. Additionally, the start and end time of each trial and the duration of overlap of the participants’ right hands were recorded.

\subsubsection{Subjective measures}
Participants filled in the Simulator Sickness Questionnaire(SSQ)~\cite{kennedy_simulator_1993} before and after the study. Additionally, participants filled in the Igroup Presence Questionnaire (IPQ)~\cite{schubert2001experience} at the end of the study. 


During Task 1 and Task 2, participants used the Oculus motion controllers to provide a Likert-scale rating ranging from "not at all" (1) to "fully in control" (7) for the question ``How much do you feel in control?'' after each trial to measure their subjective ``Sense of Agency'' over the shared avatar. These questions were embedded as panels in VR, allowing participants to stay immersed in the VR experience~\cite{Putze2020}. After each haptic feedback condition, participants would answer three questions about their ``sense of co-presence'' and three questions about their ``sense of body ownership'', taken from standard questionnaires of co-presence~\cite{pimentel_copresence_2021} and avatar embodiment~\cite{peck_avatar_2021}. Given these questions belong to two different questionnaires, we calculate the reliability scores separately. These six questions were selected based on their relevance to the study design, while reducing participants' workload and the total session time compared to using the full questionnaires.

\noindent \textbf{Co-presence (CP) questionnaire (Cronbach's $\alpha$=0.87)}
\begin{enumerate}
    \item I felt that I was in the presence of the other person
    \item I felt that the other person and I were together in the same space
    \item I felt that the other person responded to shifts in my movement (e.g., posture, position)
\end{enumerate}

\noindent \textbf{Body Ownership (BO) questionnaire (Cronbach's $\alpha$=0.58)}
\begin{enumerate}
    \item I felt as if my (real) hands were drifting toward the virtual hands or as if the virtual hands were drifting toward my (real) hands
    \item I felt as if the movements of the virtual hands were influencing my own movements
    \item At some point, it felt as if my real hands were starting to take on the posture or shape of the virtual hands that I saw
\end{enumerate}

\subsection{Hardware and software setup}
Participants performed the study using Oculus Quest 2 Head-Mounted Displays (HMDs) and Oculus Touch VR motion controllers connected to desktop computers. These computers ran the virtual environment we created using Unreal Engine 5.1\footnote{\url{https://www.unrealengine.com/}}, and were connected with Ethernet to ensure minimum latency between the computers. One computer hosts a local server while the second computer joined this server as a client. Each computer recorded the rotation and position of their respective users. To create a co-embodied avatar, the level spawns a ``shared hands'' avatar in the virtual world. This virtual representation was chosen to model a gender neutral representation of hands (cf., \cite{Schwind2017}). Since the controller was not represented in virtual space, we did not have the position of the hand to be wrapped around the controller, and instead showed a default open palm position pose. To determine the position of each of the shared hands, the avatar linearly interpolates between the position of User 1 and User 2, expressed by the following equation:

\begin{equation}
    \textit{x\textsubscript{fusion}} = \alpha\textit{x\textsubscript{user1}} + (1-\alpha)\textit{x\textsubscript{user2}} \hspace{2cm}   (0 < \alpha < 1)
    \label{ctrldist}
\end{equation}
where $\alpha$ controls the interpolation such that the resulting position is 100\% of Player 1 when $\alpha$ is 1 and 100\% of Player 2’s position when $\alpha$ is 0. This value can be set to vary the control over the shared hands in each level to W25/W50/W75 to create the conditions outlined in the study design.

\begin{figure}
    \centering
    \includegraphics[width=1\linewidth]{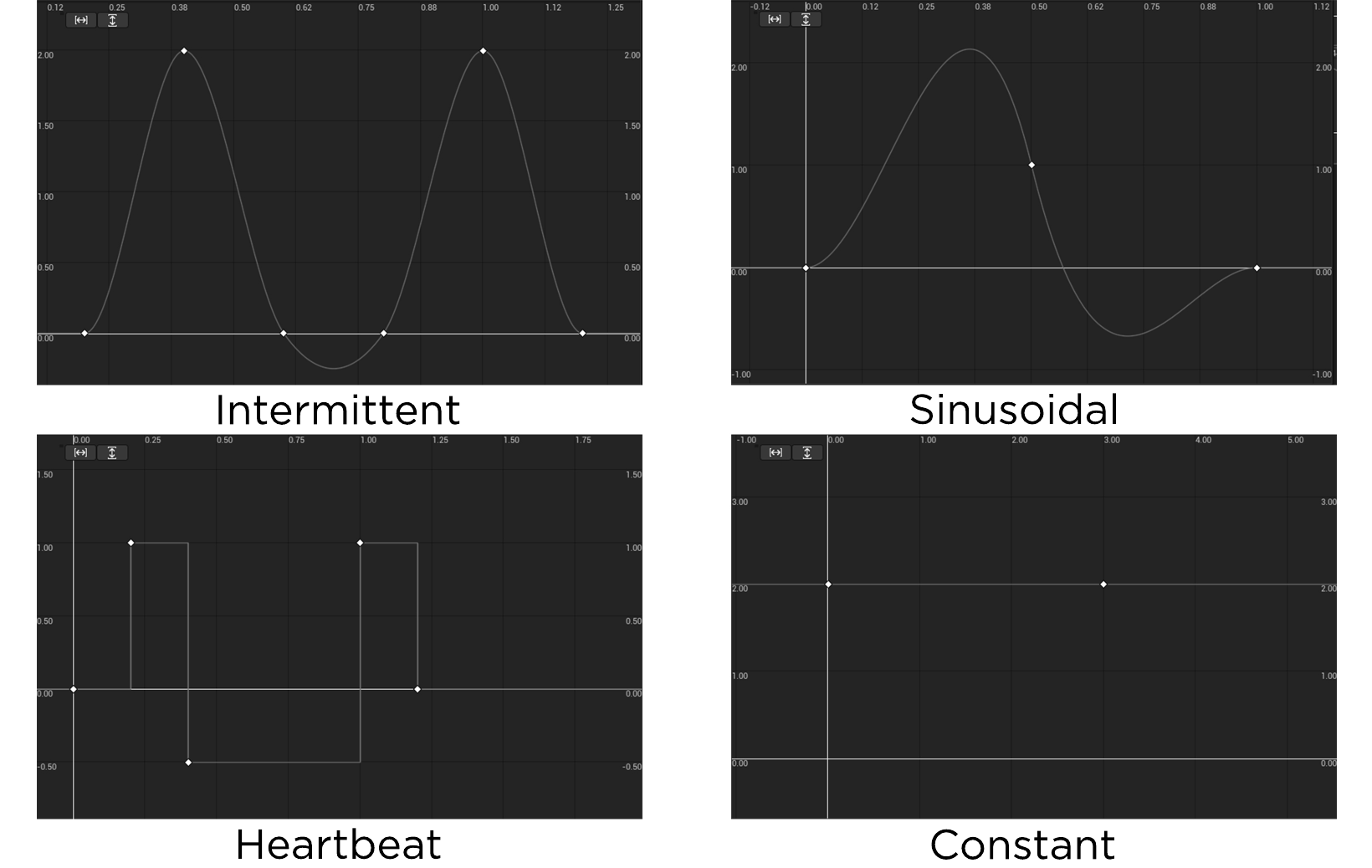}
    \caption{Waveforms of four vibration patterns used in pre-study (created in Unreal Engine): Intermittent (top left), Sinusoidal (top right), Heartbeat (bottom left) and Constant (bottom right)}
    \label{fig:waveforms}
    \Description{Four images showing the waveforms that were created in Unreal Engine for the vibration patterns; On the top left is the waveform for the intermittent pattern, and bottom left is the waveform for the Heartbeat pattern; On the top right is the waveform for the sinusoidal pattern and bottom right is the waveform for the constant pattern.}
\end{figure}

\subsubsection{Haptic feedback design}
Previous experiments by~\citet{wentzel_improving_2020} tested techniques to modulate amplification levels of vibrations and found that it impacted the user’s comfort. Since our study is modelled on the perceptual crossing paradigm~\cite{auvray_perceptual_2009}, which uses simplistic ON/OFF feedback for interactions between the participants, we also implement the haptic feedback to ON when participants hands are in the same virtual position and OFF otherwise. No additional hardware is implemented for sophisticated vibrotactile feedback cues as the scope of this study is limited to using the standard oculus motion controllers. Therefore, we conducted a pre-study to make an informed decision on the intensity and pattern of the vibrations. Fifteen Participants (Mean age = 25.66, SD = 2.09) tested four common type vibration patterns~\cite{seifi2015vibviz}: Intermittent, Sinusoidal, Heartbeat, and Constant (Figure~\ref{fig:waveforms}) in combination with four intensity levels: 10, 20, 30, and 40. The intermittent pattern consisted of two vibration occurrences every second, the sinusoidal had one occurrence per second, the heartbeat pattern consisted of two short occurrences followed by a pause every second, and the constant pattern had a continuous vibration throughout. These were tested in VR using the same motion controllers that would be used for the main study. While receiving haptic feedback, participants rated on a Likert scale their perceived comfort ranging from very uncomfortable (1) to very comfortable (7), and their perceived intensity level ranging from very calm (1) to very intense (7).

\begin{table}[t]
    \small
    \centering
    \begin{tabular}{lp{1.2cm}p{1.5cm}p{1.5cm}}
    \hline
    \textbf{Pattern type} & \textbf{Pattern intensity} & \textbf{Perceived (mean) intensity} & \textbf{Perceived (mean) comfort} \\
    \hline
    Intermittent & 10 & 3.2 & 4.4 \\ 
    Intermittent & 20 & 3.6 & 4.4 \\ 
    Intermittent & 30 & 4.5 & 3.8 \\ 
    Intermittent & 40 & 5.2 & 3.4 \\ 
    \hline
    Heartbeat & 10 & 1.8 & 5.2 \\ 
    Heartbeat & 20 & 3.3 & 4.2 \\
    Heartbeat & 30 & 3.4 & 3.9 \\
    Heartbeat & 40 & 4 & 3.8 \\
    \hline
    Sinusoidal & 10 & 3.4 & 5.1 \\
    \textbf{Sinusoidal} & \textbf{20} & \textbf{4.2} & \textbf{4.4} \\
    Sinusoidal & 30 & 5.4 & 3.8 \\
    Sinusoidal & 40 & 4.2 & 3.6 \\
    \hline
    Constant & 10 & 3.8 & 4.1 \\
    Constant & 20 & 5.3 & 3.2 \\
    Constant & 30 & 6.6 & 2.4 \\
    Constant & 40 & 6.7 & 2.2 \\
    \hline
    \end{tabular}
    \caption{Results of our pre-study on perceived mean intensity and comfort for 16 haptic patterns. The mean perceived intensity ratings range between 1.8 to 6.7 (where 1 is very calm and 7 very intense) and the mean perceived comfort ratings range between 2.2 to 5.2 (where 1 is very uncomfortable and 7 very comfortable). The chosen variant "sinusoidal pattern with intensity 20" received a mean intensity rating of 4.2 and comfort rating of 4.4.}
    \Description{A table showing the results of our pre-study on perceived mean intensity and comfort for 16 vibrotactile pattern variations, with the chosen variant for the main study highlighted. The perceived intensity ratings range between 1.8 to 6.7 and the perceived comfort ratings range between 2.2 to 5.2. The chosen variant "sinusoidal pattern with intensity 20" received a mean intensity rating of 4.2 and comfort rating of 4.4.}
    \label{tab:prestudy}
\end{table}

From Table~\ref{tab:prestudy}, results showed that high-intensity vibrations (30,40) had lower comfort ratings overall. The highest comfort rating was given to the heartbeat and sinusoidal patterns at intensity 10. However, this intensity level was often hardly noticeable by some participants. The variant with the sinusoidal pattern at intensity 20, provided a balanced level of intensity (mean score = 4.266) while still being comfortable (mean score = 4.4). Therefore, we chose to implement this variant of the haptic feedback for our main study.

\subsection{Study procedure}
\begin{figure}
    \centering
    \includegraphics[width=1\linewidth]{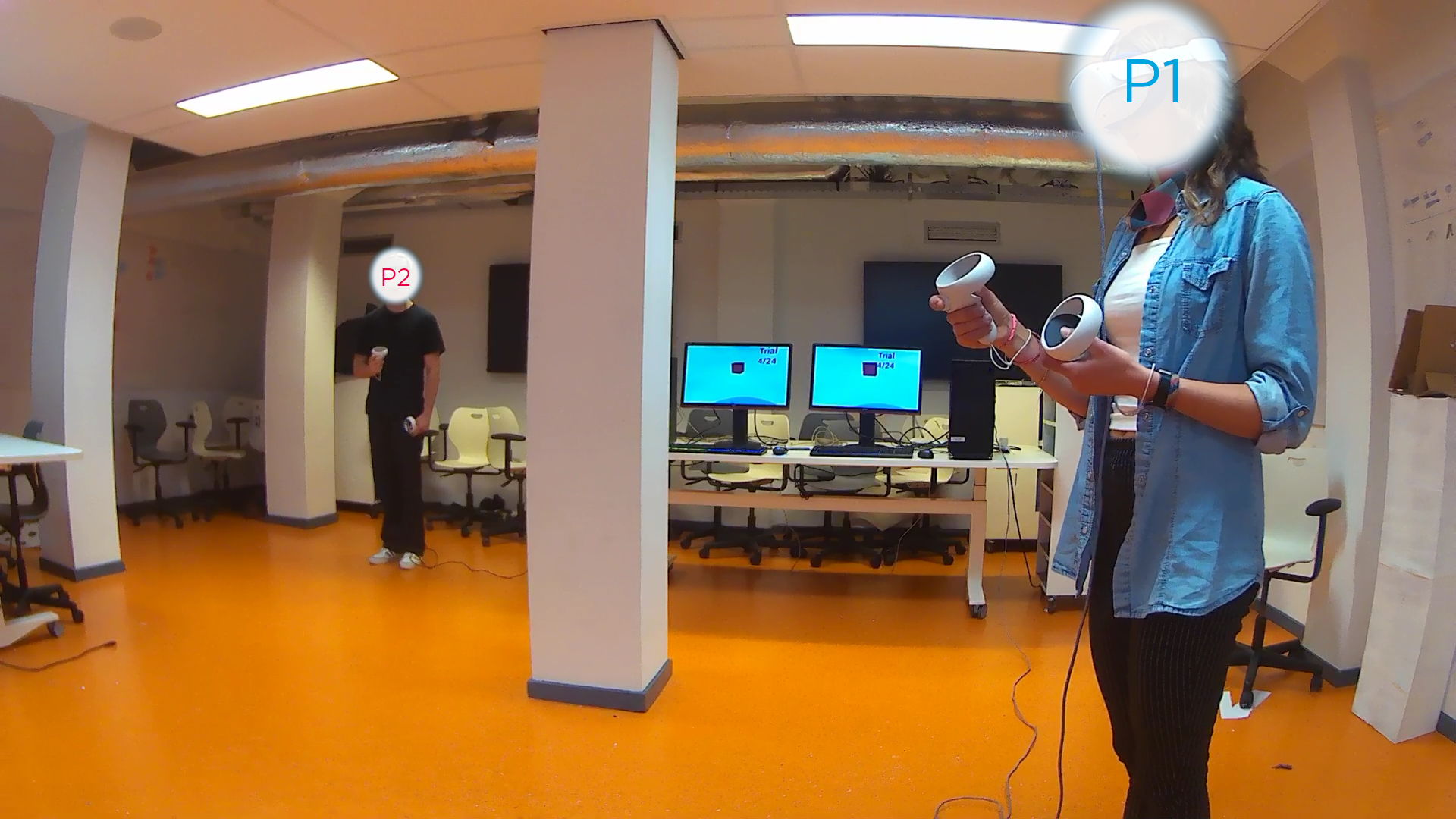}
    \caption{An image of the study setup that shows two users performing a task, and the two computer screens showing the perspective view of each user}
    \label{fig:studysetup}
    \Description{An image of the experimental setup with user 1 on the left and user 2 on the right performing the task with the computers placed on a table showing the first-person perspective of each user in the background}
\end{figure}

At the study location, a table was placed where participants would fill out all the forms: demographics, informed consent, pre- and post-study SSQ, and IPQ. Two computers were placed side by side on a separate table and were connected to HMDs (Figure~\ref{fig:studysetup}). One participant was randomly assigned to the computer acting as User 1, and the other to the computer acting as User 2. A video camera was also in place to record both participants’ motions while they performed the tasks. A video showing this interaction is provided in \textcolor{purple}{Supplementary Material A}. The position in which both participants would stand was marked on the floor. To reduce any possibility of injury and to simplify the interactions, participants conducted the trials while standing, and only used their right hand for the motion task. The spatial chime sound was channeled through the HMD speakers, set at a comfortable 60\% volume.

Upon their arrival, participants were asked to read and sign the informed consent form and fill in a pre-study SSQ. Similar to~\citet{fribourg_virtual_2021}, participants were briefed that they would be sharing the avatar during all trials and instructed to avoid communicating with each other verbally. No instructions were provided regarding the haptic feedback, allowing participants to interpret the meaning of the vibrotactile cue when it occurs during the tasks. This was done to evaluate the effectiveness of the chosen vibration pattern in establishing synchronization patterns between participants autonomously, as was done in the original perceptual crossing experiments~\cite{auvray_perceptual_2009,auvray2012perceptual}.
Participants provided subjective ratings on their sense of agency, co-presence, and body ownership after each set of trials using questionnaires presented inside the virtual environment. Each participant performed 25 trials (including repetitions), and answered 32 questions during the study (24 Sense of Agency + 4 Co-presence + 4 Body ownership). At the end of the study, participants filled in the post-study SSQ along with the IPQ. Finally, a semi-structured interview was conducted with both participants, which lasted around 15 minutes. During the interviews, we asked participants about their overall impression of the study, their perceptions of the shared motion and the haptic feedback, their impressions regarding the two tasks, and provocations regarding further use cases of virtual co-embodiment. The complete interview guide is provided in \textcolor{purple}{Supplementary Material B}. Sessions lasted an average of 60 minutes, whereas the within-VR portion took approximately 25 minutes. Each participant was compensated with a \texteuro/\$10 gift voucher for participating. Our study received approval from our institute's ethics and data protection committee, where we also followed any guidelines pertaining to any prevailing cleanliness (cf., COVID-19) regulations.  

\subsubsection{Participants}

Twenty pairs of participants\footnote{For effect size f=0.25 under $\alpha$=0.05 and power (1-$\beta$)=0.95, with 24 repeated measurements within factors, a minimum of 12 participants is needed.} (40 people, 23f, 17m) were recruited (M=25.95 years, SD=2.59). Participants were recruited primarily from the first author's university. All were right-handed. Fifteen participants reported no prior VR experience; 17 reported being novice users (having used VR at least once), and eight reported occasional VR use. Of the 20 pairs, three were couples, 12 were friends, and five did not know each other (i.e., strangers). There were three male-male pairs (two friend pairs, one stranger pair). There were six female-female pairs (five friend pairs, one stranger pair). There were eleven male-female mixed pairs (three were couples, six were friend pairs, two were stranger pairs).

\section{Analysis and results}
A mixed-methods approach was adopted for analysis, which means the results of the quantitative analysis are interpreted along with the qualitative analysis to explain the phenomena observed.

\subsection{Pre-processing and analysis approach}

\subsubsection{Synchrony measure}
Users' hand motion data was re-sampled to 100Hz. We cleaned the data by removing missing values, NAs (Not Available) due to logging errors, and duplicates. This resulted in the removal of 11,679 records, with a final dataset size = 880,549 records. Several measures of inter-personal synchrony exist, from dyadic synchrony in VR as was done by Sun et al. (2019)~\cite{Sun2019}, to breathing synchrony as was done by El Ali et al. (2023)~\cite{Elali2023BreathWithMe}. Given our dataset, we analyze joint motion synchrony by adapting Sun et al.'s~\cite{Sun2019} approach — we perform the following steps to obtain our synchrony measures: The extracted <X,Y,Z> positional movement data was used to calculate the distance moved between each timestamp for each participant. We calculate the Euclidean distances for the movement of both participants, by taking the square of the difference between the consecutive positions in each direction (X, Y and Z). We then compute the square root of the sum of these squared differences to calculate the overall Euclidean distance for the movement between each timestamp. The intervals of the timestamps that are considered for this calculation are short (in the order of milliseconds), therefore any repeated movements (left and right) that occur over an interval will be captured, and will be different from a continuous motion in a single direction.

We then compute the rolling Spearman correlation between each participants' summed (right hand) Euclidean movement. Since our Euclidean measures were not normally distributed, we used rolling Spearman's Rank Correlation Coefficient with a window size of 450 samples to compare the two movement series. The mean of the rolling Spearman's Rank Correlation Coefficient was then calculated for all 24 trials across the 20 sessions\footnote{We additionally tested cosine similarity measures, given~\citet{wohltjen2023synchrony}’s approach that used Dynamic Time Warping to calculate cosine similarity scores; however, the results were similar to ours, and therefore we only report the Spearman Rank correlation results.}.

\subsubsection{Statistical analysis approach}
The combined effects of task, control, and haptics on participants' subjective ratings of perceived Sense of Agency (SoA), co-presence, body ownership, and mean Spearman's Rank Correlation Coefficient were analyzed by fitting a full mixed-effects model for each dataset. First, the normality of the data was tested using the Shapiro-Wilk test. Results for all dependent variables showed that the data distribution significantly deviated from normality (p < 0.05). Therefore, aligned rank transforms were applied to the data before fitting it to the model~\cite{wobbrock_aligned_2011}. Holm-Bonferroni corrections were applied to the datasets, and contrast tests were conducted using ART-C~\cite{elkin_aligned_2021}. The results of the analysis of variance for all response variables are provided in Table~\ref{table:variance}.

\subsection{Quantitative results}

\begin{table*}[p]
\small
\centering
\begin{tabular}{p{3.5cm}rrrrrrr>{\raggedleft}p{1cm}r}
  \hline
\textbf{Response Variable} & \textbf{Factor} & \textbf{Level} & \textbf{Mean} & \textbf{Median} & \textbf{SD} & \textbf{\textit{F}} & \textbf{\textit{df}} & \textbf{\textit{p}} & \textbf{$\eta_{p}^{2}$}  \\ 
  \hline
Sense of Agency & \multirow{2}{*}{\raggedright \textbf{Task}} & Task 1 & 4.53 & 5.00 & 1.44 & \multirow{2}{*}{\raggedright 172.02} & \multirow{2}{*}{\raggedright 1} &  \multirow{2}{*}{\raggedright < .000***} & \multirow{2}{*}{\raggedright 0.16}  \\ 
& & Task 2 & 3.45 & 3.00 & 1.45  \\
& \multirow{2}{*}{\raggedright \textbf{Haptics}} & On & 3.86 & 4.00 & 1.57 & \multirow{2}{*}{\raggedright 12.93} & \multirow{2}{*}{\raggedright 1} & \multirow{2}{*}{\raggedright < .000***} & \multirow{2}{*}{\raggedright 0.01} \\ 
& & Off & 4.12 & 4.00 & 1.5  \\ 
& \multirow{3}{*}{\raggedright\textbf{Control}} &  W25 & 4.05 & 4.00 & 1.53 & \multirow{3}{*}{\raggedright 0.24} & \multirow{3}{*}{\raggedright 2} & \multirow{3}{*}{\raggedright 0.78\hphantom{**}} & \multirow{2}{*}{\raggedright 0}  \\ 
& & W50 & 3.97 & 4.00 & 1.54 \\
& & W75 & 3.96 & 4.00 & 1.56 \\
&\textbf{Task x Haptics} & - & - & - & - & 13.61 & 1 &  < .000*** & 0.01 \\ 
&\textbf{Task x Control} & - & - & - & - & 0.30 & 2 &  0.74\hphantom{**} & 0 \\ 
&\textbf{Haptics x Control} & - & - & - & - & 0.52 & 2 &  0.59\hphantom{**} & 0 \\ 
&\textbf{Task x Haptics x Control} & - & - & - & - & 0.05 & 2 &  0.95\hphantom{**} & 0 \\ 
  \hline
Co-presence 1 & \multirow{2}{*}{\raggedright \textbf{Task}} & Task 1 & 4.30 & 5.00 & 2.00 & \multirow{2}{*}{\raggedright 26.35} & \multirow{2}{*}{\raggedright 1} & \multirow{2}{*}{\raggedright < .000***} & \multirow{2}{*}{\raggedright 0.18}  \\ 
& & Task 2 & 5.35 & 6.00 & 1.64 \\
& \multirow{2}{*}{\raggedright \textbf{Haptics}} & On & 4.84 & 5.00 & 1.88 & \multirow{2}{*}{\raggedright 0.24} & \multirow{2}{*}{\raggedright 1} & \multirow{2}{*}{\raggedright 0.62\hphantom{**}} & \multirow{2}{*}{\raggedright 0}  \\ 
& & Off & 4.81 & 5.00 & 1.93 \\
&\textbf{Task x Haptics} & - & - & - & - &  0.80 & 1 &  0.37\hphantom{**} & 0.01 \\ 
  \hline
Co-presence 2 & \multirow{2}{*}{\raggedright \textbf{Task}} & Task 1 & 3.54
& 3.00 & 1.92 & \multirow{2}{*}{\raggedright 34.38} & \multirow{2}{*}{\raggedright 1} &  \multirow{2}{*}{\raggedright < .000***} & \multirow{2}{*}{\raggedright 0.23}  \\
& & Task 2 & 4.71 & 5.00 & 1.81 \\
& \multirow{2}{*}{\raggedright \textbf{Haptics}} & On & 4.15 & 4.00 & 1.95 & \multirow{2}{*}{\raggedright 0.03} & \multirow{2}{*}{\raggedright 1} & \multirow{2}{*}{\raggedright 0.86\hphantom{**}} & \multirow{2}{*}{\raggedright 0}  \\ 
& & Off & 4.10 & 4.00 & 1.96 \\
&\textbf{Task x Haptics} & - & - & - & - &  2.31 & 1 &  0.13\hphantom{**} & 0.02 \\ 
  \hline
Co-presence 3 & \multirow{2}{*}{\raggedright \textbf{Task}} & Task 1 & 3.68 & 4.00 & 1.80 & \multirow{2}{*}{\raggedright 28.28} & \multirow{2}{*}{\raggedright 1} &  \multirow{2}{*}{\raggedright < .000***} & \multirow{2}{*}{\raggedright 0.19}  \\ 
& & Task 2 & 4.80 & 5.00 & 1.76 \\
& \multirow{2}{*}{\raggedright \textbf{Haptics}} & On & 4.09 & 4.00 & 1.86 & \multirow{2}{*}{\raggedright 1.24} & \multirow{2}{*}{\raggedright 1} & \multirow{2}{*}{\raggedright 0.27\hphantom{**}} & \multirow{2}{*}{\raggedright 0.01}  \\ 
& & Off & 4.39 & 5.00 & 1.86 \\
&\textbf{Task x Haptics} & - & - & - & - &  0.51 & 1 &  0.48\hphantom{**} & 0 \\ 
  \hline
Body Ownership 1 & \multirow{2}{*}{\raggedright \textbf{Task}} & Task 1 & 5.29 & 5.00 & 1.08 & \multirow{2}{*}{\raggedright 1.72} & \multirow{2}{*}{\raggedright 1} & \multirow{2}{*}{\raggedright 0.19\hphantom{**}} & \multirow{2}{*}{\raggedright 0.01}  \\ 
& & Task 2 & 4.93 & 5.00 & 1.45 \\
& \multirow{2}{*}{\raggedright \textbf{Haptics}} & On & 5.00 & 5.00 & 1.27 &  \multirow{2}{*}{\raggedright 1.42} & \multirow{2}{*}{\raggedright 1} & \multirow{2}{*}{\raggedright 0.24\hphantom{**}} & \multirow{2}{*}{\raggedright 0.01}  \\ 
& & Off & 5.21 & 5.00 & 1.30 \\
&\textbf{Task x Haptics} & - & - & - & - &  0.22 & 1 &  0.64\hphantom{**} & 0 \\ 
  \hline
Body Ownership 2 & \multirow{2}{*}{\raggedright \textbf{Task}} & Task 1 & 5.15 & 5.00 & 1.30 & \multirow{2}{*}{\raggedright 0.38} & \multirow{2}{*}{\raggedright 1} &  < \multirow{2}{*}{\raggedright .001**\hphantom{*}} & \multirow{2}{*}{\raggedright 0.08}  \\
& & Task 2 & 5.55 & 6.00 & 1.47 \\
& \multirow{2}{*}{\raggedright \textbf{Haptics}} & On & 5.30 & 6.00 & 1.30 & \multirow{2}{*}{\raggedright 1.00} & \multirow{2}{*}{\raggedright 1} & \multirow{2}{*}{\raggedright 0.32\hphantom{**}} & \multirow{2}{*}{\raggedright 0.01}  \\ 
& & Off & 5.40 & 6.00 & 1.51 \\
&\textbf{Task x Haptics} & - & - & - & - &  1.31 & 1 &  0.26\hphantom{**} & 0.01 \\ 
  \hline
Body Ownership 3 & \multirow{2}{*}{\raggedright \textbf{Task}} & Task 1 & 4.43 & 4.50 & 1.40 & \multirow{2}{*}{\raggedright 0.99} & \multirow{2}{*}{\raggedright 1} &  \multirow{2}{*}{\raggedright 0.32\hphantom{**}} & \multirow{2}{*}{\raggedright 0.01}  \\ 
& & Task 2 & 4.23 & 4.00 & 1.58 \\
& \multirow{2}{*}{\raggedright \textbf{Haptics}} & On & 4.31 & 4.00 & 1.45 & \multirow{2}{*}{\raggedright 0.01} & \multirow{2}{*}{\raggedright 1} & \multirow{2}{*}{\raggedright 0.94\hphantom{**}} & \multirow{2}{*}{\raggedright 0.00}  \\ 
& & Off & 4.34 & 4.50 & 1.54 \\
&\textbf{Task x Haptics} & - & - & - & - &  0.09 & 1 &  0.77\hphantom{**} & 0 \\ 
  \hline
  Mean Rolling Spearman's Rank Correlation coefficient & \multirow{2}{*}{\raggedright \textbf{Task}} & Task 1 & 0.39 & 0.41 & 0.21 & \multirow{2}{*}{\raggedright 57.15} & \multirow{2}{*}{\raggedright 1} &  \multirow{2}{*}{\raggedright < .000***} & \multirow{2}{*}{\raggedright 0.11}  \\
  & & Task 2 & 0.27 & 0.25 & 0.20 \\
& \multirow{2}{*}{\raggedright \textbf{Haptics}} & On & 0.34 & 0.36 & 0.20 & \multirow{2}{*}{\raggedright 2.02} & \multirow{2}{*}{\raggedright 1} &  \multirow{2}{*}{\raggedright 0.15\hphantom{**}} & \multirow{2}{*}{\raggedright 0}  \\ 
& & Off & 0.32 & 0.33 & 0.22 \\
& \multirow{3}{*}{\raggedright \textbf{Control}} & W25 & 0.34 & 0.37 & 0.21 & \multirow{3}{*}{\raggedright 2.96} & \multirow{3}{*}{\raggedright 1} &  \multirow{3}{*}{\raggedright 0.08\hphantom{**}} & \multirow{3}{*}{\raggedright 0}  \\ 
& & W50 & 0.31 & 0.28 & 0.21 \\
& & W75 & 0.34 & 0.36 & 0.22 \\
&\textbf{Task x Haptics} & - & - & - & - &  1.10 & 1 &   0.29\hphantom{**} & 0 \\ 
&\textbf{Task x Control} & - & - & - & - &  0.04 & 1 &  0.83\hphantom{**} & 0 \\ 
&\textbf{Haptics x Control} & - & - & - & - &  0.25 & 1 &  0.61\hphantom{**} & 0 \\ 
&\textbf{Task x Haptics x Control} & - & - & - & - &  2.36 & 1 &  0.12\hphantom{**} & 0 \\ 
\hline
\end{tabular}
\caption{Analysis of Deviance on the full mixed-effects model for Sense of Angency (SoA), co-presence, body ownership and mean rolling Spearman Rank Correlation using Aligned Rank Transformed data. For Sense of Agency, the model shows significance for the factors Task, Haptics and the interaction of Task and Haptics (***p<0.001). For Co-presence 1, Co-presence 2 and Co-presence 3, the model shows significance only for the Task factor (***p<0.001). For Body ownership 2, the model shows significance only for the Task factor (**p<0.01) and for Mean rolling Spearman's rank correlation coefficient, the model shows significance only for the Task factor (***p<0.001).}
\Description{A table showing our analysis of deviance results on the full mixed-effects model for sense of agency (SoA), co-presence, body ownership and mean rolling Spearman Rank Correlation using Aligned Rank Transformed data. For Sense of Agency, the model shows significance for the factors Task, Haptics and the interaction of Task and Haptics. For Co-presence 1, Co-presence 2 and Co-presence 3, the model shows significance only for the Task factor. For Body ownership 2, the model shows significance only for the Task factor and for Mean rolling Spearman's rank correlation coefficient, the model shows significance only for the Task factor.}
\label{table:variance}
\end{table*}

\subsubsection{{Sense of Agency}}
%

\begin{figure}[t]
	\centering
	\subfigure[Perceived Sense of Agency (SoA) ratings for Task-Haptic conditions where lines with asterisks indicating pairwise Holm-Bonferroni-corrected significance]{\label{fig:perceivedfoc}\includegraphics[width=0.47\linewidth]{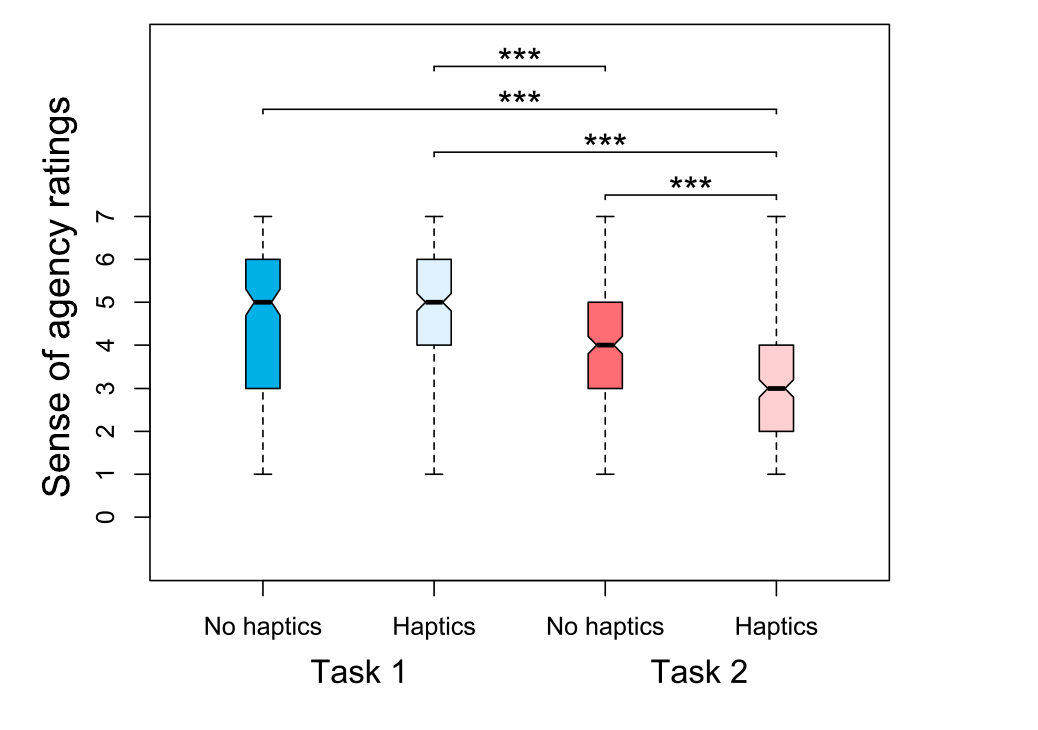}}
	\subfigure[Perceived Sense of Agency (SoA) ratings corresponding to control distribution for each task]{\label{fig:perceivedvsactualfoc}\includegraphics[width=0.47\linewidth]{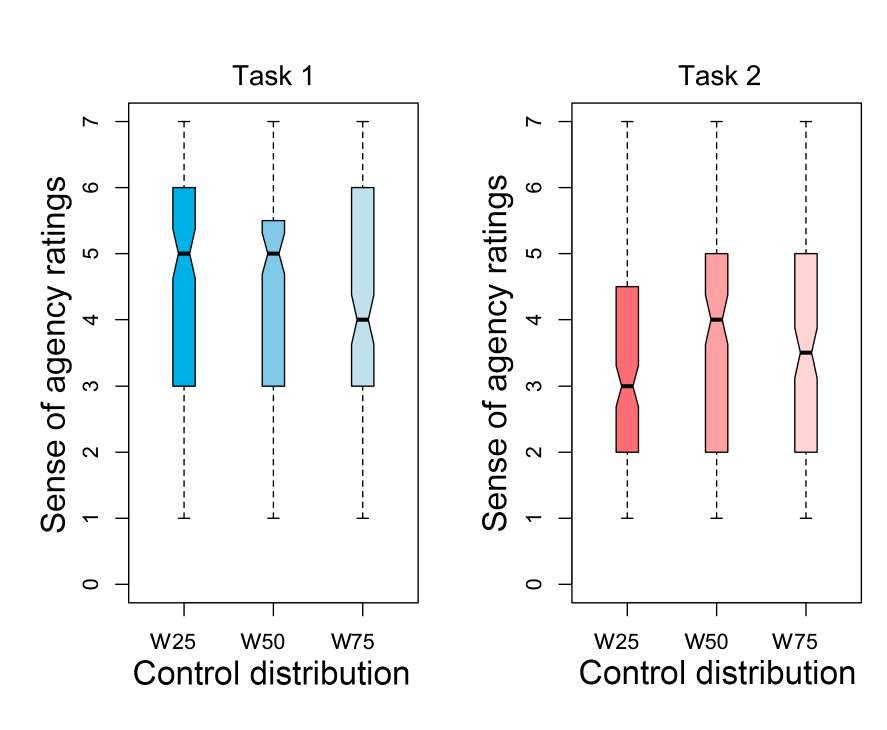}}
	\caption{Sense of agency responses to the question ``How much do you feel in control?'' asked after every trial during the study.}
	\Description{Two plots. On the left (a) the perceived sense of agency ratings per haptic feedback condition for task 1 and task 2 shown as boxplots where lines with asterisks indicate pairwise Holm-Bonferroni-corrected significance. On the right (b) the perceived sense of agency ratings versus actual control weight for task 1 and task 2 shown as boxplots.}
	\label{fig:feelingofcontrol}
\end{figure}

The analysis of the Sense of Agency (SoA) ratings are shown as boxplots in Figure~\ref{fig:perceivedfoc}, where lines with asterisks indicate pairwise, Holm-Bonferroni corrected, significance. A full mixed-effects model showed significance for Task and Haptics (p < 0.000). Significant interaction effects were also found between Task and Haptics (p < 0.000). Contrast tests for the main effect of Task revealed that responses were significantly higher in Task 1 compared to Task 2. Moreover, the contrast test for Haptics revealed that participants’ feelings of control were significantly greater in conditions without haptic feedback compared to conditions with haptic feedback. The contrast test on the interaction effects between Task and Haptics showed significant differences across all levels except between Task 1 - No haptics condition x Task 1 - Haptics condition (p = 0.219).

The comparison of the reported SoA with respect to the actual control that participants had over the shared avatar is shown as boxplots in Figure~\ref{fig:perceivedvsactualfoc}. Participants tended to overestimate and rate higher SoA when they had only 25\% control (Md=5, M=4.47, SD=1.4) and 50\% control (Md=5, M=4.49, SD=1.45) in Task 1. However, in the case of 75\% control (Md=4, M=4.48, SD=1.47) in Task 1, participants felt lower level of control over the shared avatar. We observe contrasting results for Task 2, where ratings for conditions of 25\% control (Md=3, M=3.39, SD=1.42) were rated the lowest followed by ratings for 50\% control (Md=4, M=3.45, SD=1.44). Notably, in the case of 75\% control (Md=3.5, M=3.43, SD=1.5), participants rated lower SoA in Task 2 compared to Task 1.

Given that the SoA measure was the only time series measurement we collected, we further conducted temporal analysis to assess whether these ratings changed across trials. While there were some changes in the responses (indicating participants were not randomly assigning ratings), the low correlations and lack of visible patterns did not warrant further statistical analysis. We provide this analysis (SoA rating plots over trial and correlation plot) in \textcolor{purple}{Supplementary Material C}.

\subsubsection{Co-presence}
\begin{figure}[t]
	\centering
	\subfigure[Perceived Co-presence ratings for each question for each task where lines with asterisks indicating pairwise Holm-Bonferroni-corrected significance]{\label{fig:cpratings}\includegraphics[width=0.47\linewidth]{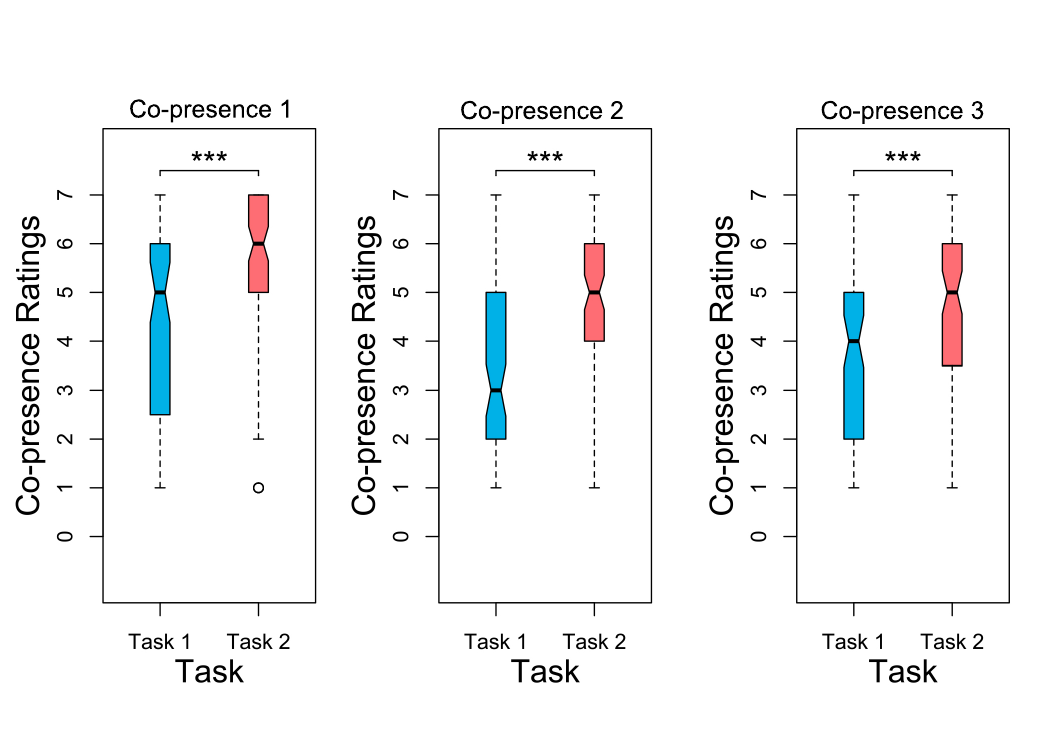}}
	\subfigure[Co-presence (CP) ratings per question for haptics conditions]{\label{fig:cphaptics}\includegraphics[width=0.47\linewidth]{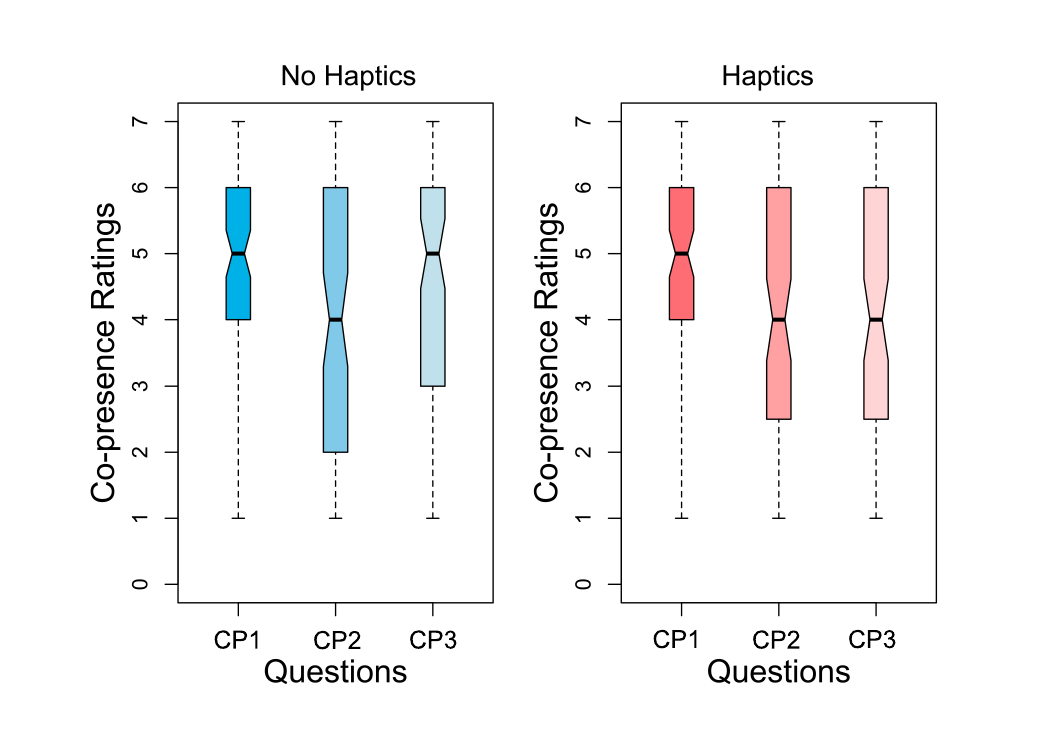}}
	\caption{Co-presence (CP) questionnaire responses. Co-presence }1: \textit{``I felt that I was in the presence of the other person"}; Co-presence 2: \textit{``I felt that the other person and I were together in the same space"}; Co-presence 3: \textit{``I felt that the other person responded to shifts in my movement"}
	\Description{Two plots. On the left (a) the perceived co-presence ratings per task for three questions of co-presence shown as boxplots where lines with asterisks indicate pairwise Holm-}Bonferroni-corrected significance. On the right (b) the perceived copresence ratings per question for haptic feedback conditions shown as boxplots.
	\label{fig:copresence}
\end{figure}

Participant ratings of the co-presence questionnaire are visualized as boxplots in Figure~\ref{fig:copresence}, where lines with asterisks indicate pairwise Holm-Bonferroni-corrected significance. A full mixed-effects model showed significance only for Task across all three responses. No significant interaction effects were found. Contrasts test showed that co-presence ratings were significantly higher in Task 2 compared with Task 1 for Co-presence 1, Co-presence 2, and Co-presence 3.

\subsubsection{Body ownership} \label{ownershipresults}
\begin{figure}[t]
	\centering
    \subfigure[Perceived body ownership (BO) ratings over virtual hands for each question for each task where lines with asterisks indicating pairwise Holm-Bonferroni-corrected significance]{\label{fig:soeratings}\includegraphics[width=0.47\linewidth]{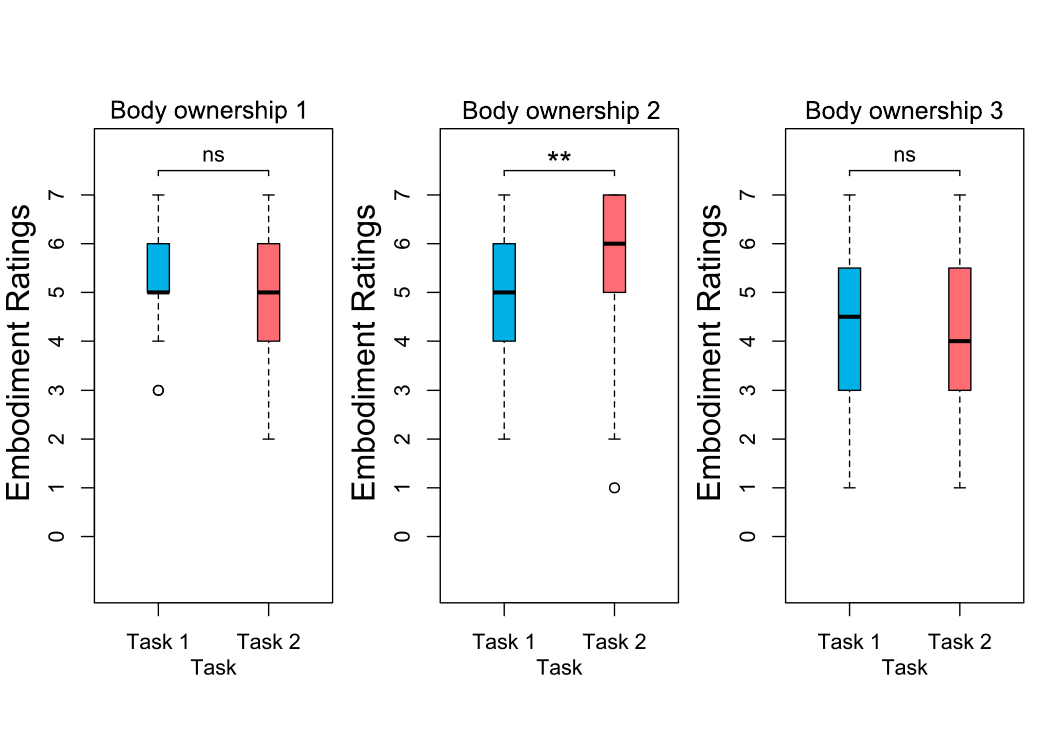}}
	\subfigure[Body ownership (BO) ratings per question for haptics conditions]{\label{fig:soehaptics}\includegraphics[width=0.47\linewidth]{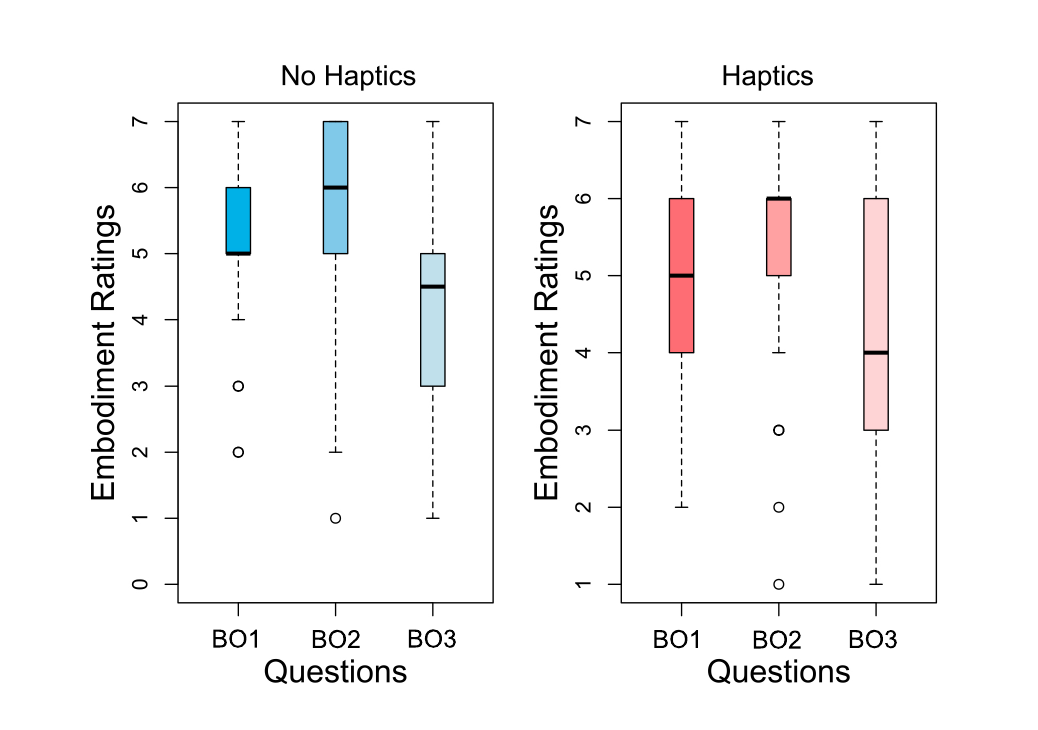}}
	\caption{Body ownership (BO) questionnaire responses. Body ownership }1: \textit{``I felt as if my (real) hands were drifting toward the virtual hands or as if the virtual hands were drifting toward my (real) hands"}; Body ownership 2: \textit{``I felt as if the movements of the virtual hands were influencing my own movements"}; Body ownership 3: \textit{``At some point, it felt as if my real hands were starting to take on the posture or shape of the virtual hands that I saw"}
	\Description{Two plots. On the left (a) the perceived body ownership ratings per task for three questions of body ownership shown as boxplots. On the right (b) the perceived body ownership ratings per question for haptic feedback conditions shown as boxplots}
	\label{fig:ownership}
\end{figure}

Analysis of participant ratings for the body ownership questionnaire are visualized as boxplots in Figure~\ref{fig:soeratings}, where lines with asterisks indicate pairwise Holm-Bonferroni-corrected significance. A full mixed-effects model showed significance only for Task for body ownership 2 responses. No significant interaction effects were found. Contrast tests showed that body Ownership 2 ratings were significantly higher in Task 2 compared with Task 1.

\subsubsection{Controller motion synchronization}
\label{sec:synch}
\begin{figure}[t]
    \centering
    \includegraphics[width=0.9\linewidth]{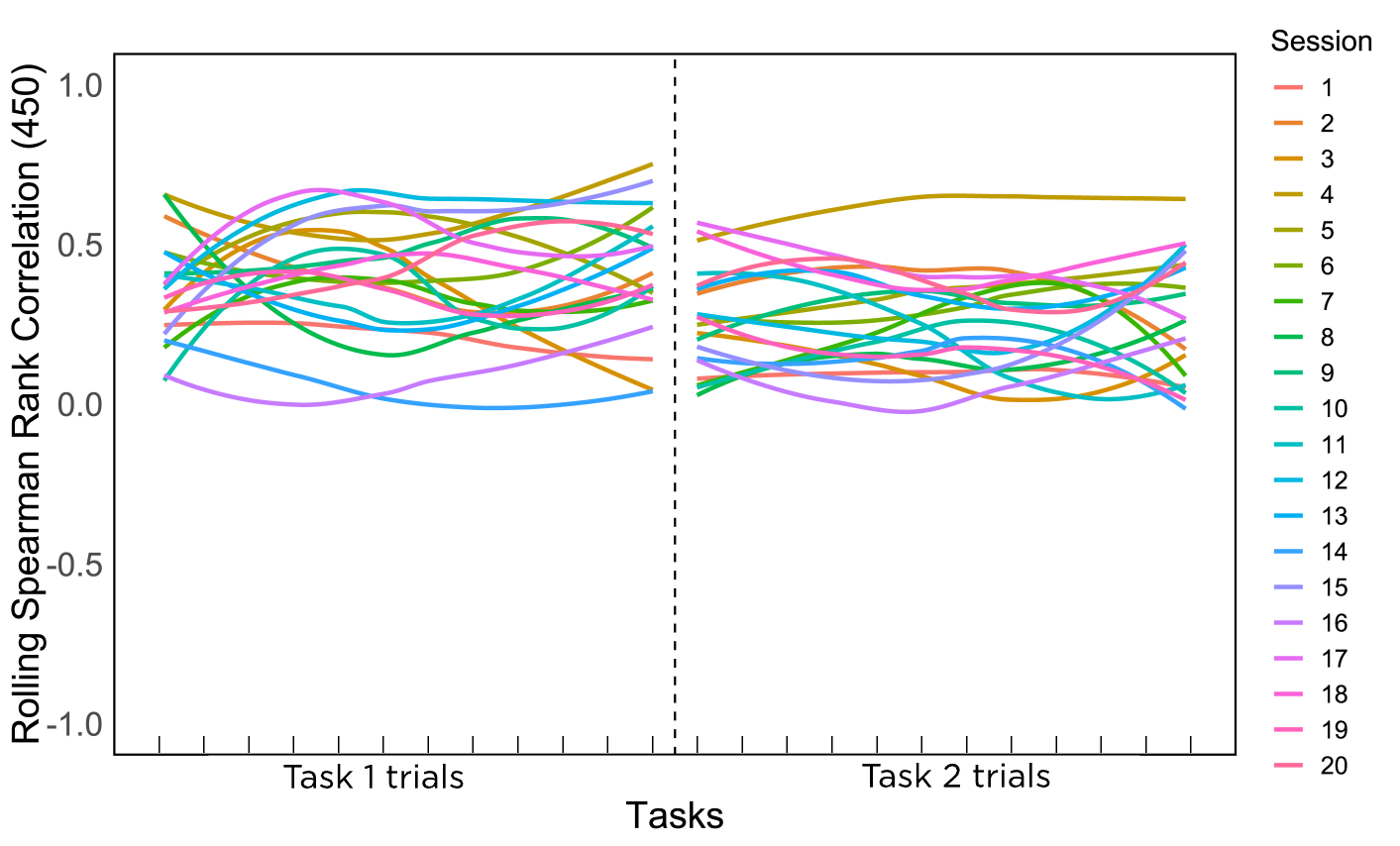}
    \caption{Motion controller synchronization (using rolling Spearman's Rank Correlation with 450 samples) between users. Plot shows synchrony across trials for each session}
    \Description{Motion controller synchronization (using rolling Spearman's Rank Correlation with 450 samples) between users. Plot shows synchrony across trials for each session.}
    \label{fig:motionsync}
\end{figure}

The analysis of participants' controller motion synchronization is visualized as time-series plot in Figure~\ref{fig:motionsync}. A full mixed-effects model showed significance for Task. No significant interaction effects were found. Contrast tests for the main effect of Task revealed that motion synchronization was significantly higher in Task 1 than Task 2. 

\subsubsection{IPQ Presence ratings}
The IPQ~\cite{schubert2001experience} has a 7-point Likert scale, ranging from -3 to 3, where this was transformed to a scale of 1 to 7 during analysis. Results with respect to each presence factor within IPQ reveals that participants experienced high levels of Involvement (M=5.08, SD=1.49) and Spatial Presence (M=4.29, SD=1.74), but felt only neutral levels of General Presence (M=3.52, SD=1.78) and Realism (M=3.52, SD=1.92).
\subsubsection{SSQ Motion sickness}
Participants’ reported motion sickness was measured before and after the study  using the SSQ questionnaire~\cite{kennedy_simulator_1993}. A Wilcoxon signed-rank test was conducted since the data did not have a normal distribution. Results showed significant differences between the pre-study (Md=1.125, IQR=0.31) and post-study (Md=1.28, IQR=0.39) scores (Z=-4.03, p<0.01, r=-0.63) indicating that participants did experience motion sickness during the study, even if slight.
\subsection{Qualitative results}

We used an inductive thematic analysis~\cite{braun2012thematic} approach. First, the lead author extensively reviewed the interview transcripts and recorded videos, generating initial codes and themes. Then, the other authors reviewed the codes and themes for consistency and offered additional themes as needed. Quotes are attributed to participants by indicating which pair (P1-P20) they belonged to, followed by the specific participant (PN-1 or PN-2) where appropriate.

\subsubsection{Frustration with shared decisions}
Participants associated their experience while performing Task 1 with being more comfortable than during Task 2, stating that ``I didn't feel much [shared control] in the beginning but in the second task with the choice it felt horrible''~[P10-2]. In fact, many participants expressed that, without a shared goal, they often ``[...] thought that I'm not controlling and somebody's here to control the hands, and it made me a bit angry'' [P15-2]. Overall, not only did participants feel more at ease with sharing the motion when a common target was presented, but the addition of free choices in Task 2 added confusion and frustration.

\subsubsection{Perception of shared motion}
Only about half of the participants (21/40) were consciously aware that the motion of the avatar was shared between them and their partner during Task 1, while the rest expressed that it only became evident to them during Task 2, when differences in choices emerged between them and their partners. Participants attributed the differences between their motion and that of the avatar as ``glitches'' or ``delays'', rather than the input of the other person in the pair. For instance, [P8-1] mentioned, ``In the beginning, it felt like the hand was not working well'', and [P16-2] remarked, ``I saw this (movement), and I thought it was an algorithm''. Importantly, during Task 2, when participants felt a diminished sense of agency or their partners’ movements were not well coordinated, they exaggerated their movements. This compensated for a perceived lack of responsiveness in shared motion: ``When I moved my hand, I noticed the hand didn’t move that much, so to compensate for it, I had to reach out more'' [P8-1]. This was particularly evident when there was a substantial difference in height between the pairs of participants, where there was an imbalance of control due to the taller participants’ extended reach, which resulted in a frustrating experience with shared motion for the partner with less reach. 


\subsubsection{Following and leading}
\label{sec:following}
Participants spontaneously took on a more \textit{follower} or \textit{leader} role during the trials. While some participants focused on actively following their partners’ movements, aiming to coordinate their actions better, others took over the lead by \textit{misbehaving} and ``[...] trying to check control by doing the opposite movement''[P3-2], so that the share of control was more apparent. During Task 2, this difference was more evident, with \textit{followers} expressing that since, ``[...]I had no control, I thought I would follow whatever pattern in movement the other person was doing'' [P11-1]. On the other hand, \textit{leaders} used this ambiguity by moving their hands dramatically to shift the shared hand, regardless of the amount of control they had over the avatar: ``I can sort of limit other persons’ actions and actually feel more in control''~[P14-1]. Additionally, five participant pairs noted that the relationship with their partners also influenced the degree of co-operation they were inclined to achieve. For example, pairs that knew each other [P7] mentioned that they would be more attentive to the other person's movement had it been with an unfamiliar person.

\subsubsection{Motion synchrony}
A high level of motion synchronization was observed during Task 1; participants started the study with distinct motions, which eventually joined when one participant began mimicking the hand motion of the other. Participants made similar observations, referring to these synchronizations as ``rhythms'' or ``flows''. For example, [P1-1] mentioned: ``after a few rounds it felt like we were getting into this rhythm,'' and [P2-2] stated: ``I started with arc motion and [P2-1] was doing a different motion, then [P2-1] started moving with arc motion''.

\subsubsection{Perception of vibration patterns and associations}\label{negvibration}

Several participants did not fully grasp during the study that haptic feedback would occur when their hands overlapped with their partner, while others inferred negative associations with the haptic feedback during the study, based on their prior experience with vibration feedback patterns. For example, participants expressed that they interpreted the haptics as hostile: ``I thought maybe I was wrong that’s why the vibrations are coming to push me in another direction'' [P6-2] or that the haptic feedback was ``[...] very random, like it was malfunctioning'' [P7-2]. Participants who viewed the haptics as positive feedback tended to associate it with video games: ``I play the Nintendo Switch, and if you win in the game, it will have vibration'' [P18-2].

\section{Discussion}
Below we discuss our study limitations and future work, and thereafter discuss our key findings by interpreting and synthesizing the results of our quantitative and qualitative analysis.

\subsection{Study limitations and future work}

First, we tested only a subset of questions in common co-presence and embodiment questionnaires -- while such additional measures could shed further light on the experience of virtual co-embodiment, this was a deliberate design choice to ensure users do not experience fatigue and overload during the study. Second, while we followed closely \citet{jeunet_you_2018}'s question regarding the sense of agency, we found that some participants may have misinterpreted what was meant by `feeling of control'. They judged the question to be related to success in the task rather than actual control over their body movements. Indeed, agency within HCI can have multiple interpretations (see \cite{Bennett2023agency} for a review), and we see this as a promising avenue for future work to explore other methods for evaluating the sense of agency in such shared virtual co-embodiment experiences. Third, it may be worthwhile to further extend the basic perceptual crossing paradigm in future work, by systematically investigating how varying the presence and type of haptics-related instructions and training beforehand would influence participants' shared agency during co-embodiment tasks. Fourth, we restricted ourselves to studying hand ownership, we do not investigate realistic full-body avatar representations (cf.,~\cite{fribourg_virtual_2021, kodama_enhancing_2022}). Furthermore, previous studies have shown that the realism of the avatar \cite{fribourg_avatar_2020} and users' choice of avatar \cite{Lim2009} impacts their sense of embodiment. Given our focus was on better understanding the role of haptic feedback and shared control distribution across targeted and free-choice tasks, we kept our study variables to a minimum to avoid blowing up the parameter space. However, this provides an interesting area for further research -- does the type of avatar body, or mixed hand representation shared amongst users similarly influences the sense of agency and co-presence? Fifth, it worth exploring how height differences between participants and their reach can impact experiences of shared avatar control. To this end, prior work has developed methods that can generate avatar body characteristics that can adapt to variable heights of participants that can be used \cite{ye_neural3points_2022} -- this would help ensure that control is distributed precisely between the participants, even if this does not necessarily reflect real-world user characteristics. Finally, given our finding that the type of relationship with another person can influence following and leading behavior (cf., Sec \ref{sec:following}), this opens up opportunities to further examine co-embodiment interactions in different dyad compositions.

\subsection{Elucidating the role of haptic feedback and avatar control distribution for virtual avatar co-embodiment}

Our study explored the impact of including haptic feedback and varying avatar control distribution on users' sense of agency, co-presence, body ownership and motion synchrony across reaching tasks in a virtual co-embodiment scenario (\textbf{RQ}). To this end, one key objective was to assess how the condition in which participants would receive haptic feedback when their hands overlapped would affect these three factors in scenarios involving shared goals and free choice. Our findings indicate that the presence of haptic feedback yielded a significant effect on the sense of agency during our study, though in unexpected ways. Participants felt a significantly greater sense of agency during conditions without haptic feedback compared to conditions with haptic feedback. Given that haptic feedback is well-suited for conveying non-verbal cues~\cite{moll2009communicative}, we expected that haptics would facilitate, not hinder sensori-motor coordination and guidance~\cite{melo2020multisensory}. Furthermore, our qualitative findings {(Sec \ref{negvibration})} indicated that the vibrotactile patterns within our haptic feedback were perceived to be a hindrance and at times intrusive. Despite that we took care to ensure pleasant vibrotactile patterns through a pre-study, as participants reported, they were reminded of smartphone and smartwatch vibrations and notifications. To interpret this finding, we first note that given our focus on translating elements of perceptual crossing into the avatar co-embodiment paradigm, we restricted our study to the context of autonomous interaction processes during shared perceptual activities. This means that even without conscious awareness of the vibrotactile cues, we expected that such position-aware haptic feedback mechanisms would support shared perceptual experiences, in this case, shared motor activity during targeted and free-choice reaching tasks. However, given the salience of the haptic stimuli, we suspect that haptics may have lowered the sense of agency for participants as they may have felt overwhelmed by the other users' guidance. This, along with the interplay of control, coordination, and physical attributes, would have then played a role in shaping the strategies the participants used in synchronizing with the other user during the haptic feedback conditions. Together, the foregoing raise cautions about how haptics can be integrated, suggesting that including vibrotactile-based haptic feedback as a positional guidance mechanism in 3D virtual space during such shared control interactions may not be an effective means for improving shared avatar co-embodiment experiences.

\subsection{Shared control, motor synchrony, and perceptual crossing across targeted and free-choice tasks}

Our findings indicate that participants' reported feelings of control (SoA) do not align with the actual levels of control. We found that participants' sense of agency increased between 25\% and 50\% control conditions, while a decrease was observed between 50\% and 75\% conditions. This result echoes the findings of Kodama et al. (2023)~\cite{kodama_effects_2023}, who did not find a clear differentiation between the tested levels of control. Moreover, we found that participants felt a significantly greater sense of agency in Task 1 (targeted) compared with Task 2 (free-choice). Since the conditions were counterbalanced and trials randomized, participants may have had a difficult time to judge absolute control levels, as they had no relative comparison to indicate such experienced control levels. We use absolute judgements for measurement of subjective responses since it is the standard practice across prior work~\cite{hagiwara_shared_2019,fribourg_virtual_2021} that investigates the sense of agency and also provided a means to test if haptics would lead to  higher (perceived) sense of agency in a given trial, without referencing back to earlier trials (which may not have had haptics activated). As such, overestimation of control was apparent during Task 1. This lends credence to the findings by~\citet{fribourg_virtual_2021}, who also found that participants perceived a greater sense of agency when the goal was shared compared to situations where participants pursued different goals. Furthermore, participants felt a significantly greater sense of co-presence during Task 2 compared with Task 1, suggesting that strong motion synchronization effects may have diminished the awareness of the other. This was further echoed by participants, where some reported a lack of awareness that they were sharing an avatar with their partners during Task 1. Qualitative analyses of user responses in the perceptual crossing paradigm also highlight that movement synchronization might not always necessarily indicate recognition of each other~\cite{kojima2017sensorimotor}. In our specific implementation, in Task 1 (targeted), it might have been the case that the straight-forward task environment afforded high synchronization, but no real space for active exploration to take place (i.e., similar to the oscillating movements found upon successful recognition of each other in perceptual crossing studies) limiting opportunities to become aware of the other participant. As such, participants' attention to their partners' movement and need to explicitly communicate verbally with their partners during Task 2 also indicates that they were more inclined to consciously co-ordinate, compared to the more autonomous interaction that was observed during Task 1. 

We also found that participants felt that the movements of the virtual hands were influencing their own movements (Body ownership 2; cf., Sec. \ref{ownershipresults}) significantly greater during Task 2 compared with Task 1. This was further reflected upon by participants who stated they actively strategized to either exert more control over the virtual hand or to follow its movements during Task 2. Interestingly, similar strategizing about movements when encountering other users are found in the perceptual crossing paradigm, where in some cases users would spontaneously adopt leader and follower roles~\cite{froese2014using}. For example, users may choose to remain stationary and passively receive the other's touch~\cite{kojima2017sensorimotor}. These parallels are interesting given the stark differences in available sensory information in our tasks compared to the perceptual crossing paradigm. It raises interesting questions about ways in which the basic paradigm can be extended and integrated into more multi-sensory shared virtual environments. The foregoing raise fundamental questions about the nature of shared control and social coordination as we integrate with machines and one another \cite{Mueller2020}: to what extent should we be consciously aware of bodily feedback mechanisms during shared activities? Given the importance of motion synchrony in varying the levels of conscious awareness of the virtual other, to what extent should shared body control systems, whether with humans or machines, leverage this without impeding on users' sense of perceived and actual agency?

\section{Conclusion}
We investigated whether integrating haptics into shared avatar co-embodiment can enhance users' shared VR experiences. Drawing on the perceptual crossing paradigm, we examine whether implementing non-verbal feedback mechanisms (namely, haptic feedback) within embodied interaction between two users can improve such social coordination experiences. Insights from this work provide a deeper understanding of the dynamics between users during co-embodiment and its impact on the perceptions of their sense of agency, co-presence, and body ownership towards a virtual hand. We found that haptic feedback given to participants when their hands overlapped led to a diminished sense of agency during co-embodiment. Our findings showed (a) a lower sense of agency in the free-choice with haptics compared to no feedback, (b) higher agency during the shared target task, (c) co-presence and embodiment were significantly higher in tasks where there were multiple targets, (d) users’ hand motions synchronized more in the targeted task. Our work contributes a deeper understanding and cautionary considerations for the role of vibrotactile haptic feedback and shared control distribution in the emerging area of virtual avatar co-embodiment.

\bibliographystyle{ACM-Reference-Format}
\interlinepenalty=10000
\bibliography{main}


\begin{thebibliography}{69}


\ifx \showCODEN    \undefined \def \showCODEN     #1{\unskip}     \fi
\ifx \showDOI      \undefined \def \showDOI       #1{#1}\fi
\ifx \showISBNx    \undefined \def \showISBNx     #1{\unskip}     \fi
\ifx \showISBNxiii \undefined \def \showISBNxiii  #1{\unskip}     \fi
\ifx \showISSN     \undefined \def \showISSN      #1{\unskip}     \fi
\ifx \showLCCN     \undefined \def \showLCCN      #1{\unskip}     \fi
\ifx \shownote     \undefined \def \shownote      #1{#1}          \fi
\ifx \showarticletitle \undefined \def \showarticletitle #1{#1}   \fi
\ifx \showURL      \undefined \def \showURL       {\relax}        \fi
\providecommand\bibfield[2]{#2}
\providecommand\bibinfo[2]{#2}
\providecommand\natexlab[1]{#1}
\providecommand\showeprint[2][]{arXiv:#2}

\bibitem[\protect\citeauthoryear{Argelaguet, Hoyet, Trico, and Lecuyer}{Argelaguet et~al\mbox{.}}{2016}]%
        {Argelaguet2016}
\bibfield{author}{\bibinfo{person}{Ferran Argelaguet}, \bibinfo{person}{Ludovic Hoyet}, \bibinfo{person}{Michael Trico}, {and} \bibinfo{person}{Anatole Lecuyer}.} \bibinfo{year}{2016}\natexlab{}.
\newblock \showarticletitle{The role of interaction in virtual embodiment: Effects of the virtual hand representation}. In \bibinfo{booktitle}{\emph{2016 IEEE Virtual Reality (VR)}}. \bibinfo{publisher}{IEEE}, \bibinfo{address}{Greenville, SC, USA}, \bibinfo{pages}{3--10}.
\newblock
\urldef\tempurl%
\url{https://doi.org/10.1109/VR.2016.7504682}
\showDOI{\tempurl}


\bibitem[\protect\citeauthoryear{Auvray, Lenay, and Stewart}{Auvray et~al\mbox{.}}{2009}]%
        {auvray_perceptual_2009}
\bibfield{author}{\bibinfo{person}{Malika Auvray}, \bibinfo{person}{Charles Lenay}, {and} \bibinfo{person}{John Stewart}.} \bibinfo{year}{2009}\natexlab{}.
\newblock \showarticletitle{Perceptual interactions in a minimalist virtual environment}.
\newblock \bibinfo{journal}{\emph{New Ideas in Psychology}} \bibinfo{volume}{27}, \bibinfo{number}{1} (\bibinfo{date}{April} \bibinfo{year}{2009}), \bibinfo{pages}{32--47}.
\newblock
\showISSN{0732118X}
\urldef\tempurl%
\url{https://doi.org/10.1016/j.newideapsych.2007.12.002}
\showDOI{\tempurl}


\bibitem[\protect\citeauthoryear{Auvray and Rohde}{Auvray and Rohde}{2012}]%
        {auvray2012perceptual}
\bibfield{author}{\bibinfo{person}{Malika Auvray} {and} \bibinfo{person}{Marieke Rohde}.} \bibinfo{year}{2012}\natexlab{}.
\newblock \showarticletitle{Perceptual crossing: the simplest online paradigm}.
\newblock \bibinfo{journal}{\emph{Frontiers in human neuroscience}}  \bibinfo{volume}{6} (\bibinfo{year}{2012}), \bibinfo{pages}{181}.
\newblock


\bibitem[\protect\citeauthoryear{Bennett, Metatla, Roudaut, and Mekler}{Bennett et~al\mbox{.}}{2023}]%
        {Bennett2023agency}
\bibfield{author}{\bibinfo{person}{Dan Bennett}, \bibinfo{person}{Oussama Metatla}, \bibinfo{person}{Anne Roudaut}, {and} \bibinfo{person}{Elisa~D. Mekler}.} \bibinfo{year}{2023}\natexlab{}.
\newblock \showarticletitle{How Does HCI Understand Human Agency and Autonomy?}. In \bibinfo{booktitle}{\emph{Proceedings of the 2023 CHI Conference on Human Factors in Computing Systems}} (Hamburg, Germany) \emph{(\bibinfo{series}{CHI '23})}. \bibinfo{publisher}{Association for Computing Machinery}, \bibinfo{address}{New York, NY, USA}, Article \bibinfo{articleno}{375}, \bibinfo{numpages}{18}~pages.
\newblock
\showISBNx{9781450394215}
\urldef\tempurl%
\url{https://doi.org/10.1145/3544548.3580651}
\showDOI{\tempurl}


\bibitem[\protect\citeauthoryear{Carvalho, Teixeira, Lucas, Yuan, Chaves, Peressutti, Machado, Bittencourt, Men{\'e}ndez-Gonz{\'a}lez, Nardi, et~al\mbox{.}}{Carvalho et~al\mbox{.}}{2013}]%
        {carvalho2013mirror}
\bibfield{author}{\bibinfo{person}{Diana Carvalho}, \bibinfo{person}{Silmar Teixeira}, \bibinfo{person}{Marina Lucas}, \bibinfo{person}{Ti-Fei Yuan}, \bibinfo{person}{Fernanda Chaves}, \bibinfo{person}{Caroline Peressutti}, \bibinfo{person}{Sergio Machado}, \bibinfo{person}{Juliana Bittencourt}, \bibinfo{person}{Manuel Men{\'e}ndez-Gonz{\'a}lez}, \bibinfo{person}{Antonio~Egidio Nardi}, {et~al\mbox{.}}} \bibinfo{year}{2013}\natexlab{}.
\newblock \showarticletitle{The mirror neuron system in post-stroke rehabilitation}.
\newblock \bibinfo{journal}{\emph{International archives of medicine}} \bibinfo{volume}{6}, \bibinfo{number}{1} (\bibinfo{year}{2013}), \bibinfo{pages}{1--7}.
\newblock


\bibitem[\protect\citeauthoryear{Cooper, Camic, Long, Panter, Rindskopf, and Sher}{Cooper et~al\mbox{.}}{2012}]%
        {braun2012thematic}
\bibfield{author}{\bibinfo{person}{Harris~Ed Cooper}, \bibinfo{person}{Paul~M Camic}, \bibinfo{person}{Debra~L Long}, \bibinfo{person}{AT Panter}, \bibinfo{person}{David~Ed Rindskopf}, {and} \bibinfo{person}{Kenneth~J Sher}.} \bibinfo{year}{2012}\natexlab{}.
\newblock \showarticletitle{Thematic analysis}. In \bibinfo{booktitle}{\emph{{APA Handbook of Research Methods in Psychology: Vol. 2. Research Designs}}}. \bibinfo{publisher}{American Psychological Association}, \bibinfo{address}{Washington}, \bibinfo{pages}{57--71}.
\newblock
\urldef\tempurl%
\url{https://doi.org/10.1037/13620-000}
\showDOI{\tempurl}


\bibitem[\protect\citeauthoryear{Cornelio, Haggard, Hornbaek, Georgiou, Bergstr{\"o}m, Subramanian, and Obrist}{Cornelio et~al\mbox{.}}{2022}]%
        {Cornelio2022}
\bibfield{author}{\bibinfo{person}{Patricia Cornelio}, \bibinfo{person}{Patrick Haggard}, \bibinfo{person}{Kasper Hornbaek}, \bibinfo{person}{Orestis Georgiou}, \bibinfo{person}{Joanna Bergstr{\"o}m}, \bibinfo{person}{Sriram Subramanian}, {and} \bibinfo{person}{Marianna Obrist}.} \bibinfo{year}{2022}\natexlab{}.
\newblock \showarticletitle{The sense of agency in emerging technologies for human-computer integration: A review}.
\newblock \bibinfo{journal}{\emph{Front. Neurosci.}}  \bibinfo{volume}{16} (\bibinfo{date}{Sept.} \bibinfo{year}{2022}), \bibinfo{pages}{949138}.
\newblock


\bibitem[\protect\citeauthoryear{El~Ali, Stepanova, Palande, Mader, Cesar, and Jansen}{El~Ali et~al\mbox{.}}{2023}]%
        {Elali2023BreathWithMe}
\bibfield{author}{\bibinfo{person}{Abdallah El~Ali}, \bibinfo{person}{Ekaterina~R. Stepanova}, \bibinfo{person}{Shalvi Palande}, \bibinfo{person}{Angelika Mader}, \bibinfo{person}{Pablo Cesar}, {and} \bibinfo{person}{Kaspar Jansen}.} \bibinfo{year}{2023}\natexlab{}.
\newblock \showarticletitle{BreatheWithMe: Exploring Visual and Vibrotactile Displays for Social Breath Awareness during Colocated, Collaborative Tasks}. In \bibinfo{booktitle}{\emph{Extended Abstracts of the 2023 CHI Conference on Human Factors in Computing Systems}} (Hamburg, Germany) \emph{(\bibinfo{series}{CHI EA '23})}. \bibinfo{publisher}{Association for Computing Machinery}, \bibinfo{address}{New York, NY, USA}, Article \bibinfo{articleno}{58}, \bibinfo{numpages}{8}~pages.
\newblock
\showISBNx{9781450394222}
\urldef\tempurl%
\url{https://doi.org/10.1145/3544549.3585589}
\showDOI{\tempurl}


\bibitem[\protect\citeauthoryear{Elkin, Kay, Higgins, and Wobbrock}{Elkin et~al\mbox{.}}{2021}]%
        {elkin_aligned_2021}
\bibfield{author}{\bibinfo{person}{Lisa~A. Elkin}, \bibinfo{person}{Matthew Kay}, \bibinfo{person}{James~J. Higgins}, {and} \bibinfo{person}{Jacob~O. Wobbrock}.} \bibinfo{year}{2021}\natexlab{}.
\newblock \bibinfo{title}{An {Aligned} {Rank} {Transform} {Procedure} for {Multifactor} {Contrast} {Tests}}.
\newblock
\newblock
\urldef\tempurl%
\url{https://doi.org/10.48550/arXiv.2102.11824}
\showDOI{\tempurl}
\newblock
\shownote{arXiv:2102.11824 [cs, stat].}


\bibitem[\protect\citeauthoryear{Fribourg, Argelaguet, Lecuyer, and Hoyet}{Fribourg et~al\mbox{.}}{2020}]%
        {fribourg_avatar_2020}
\bibfield{author}{\bibinfo{person}{Rebecca Fribourg}, \bibinfo{person}{Ferran Argelaguet}, \bibinfo{person}{Anatole Lecuyer}, {and} \bibinfo{person}{Ludovic Hoyet}.} \bibinfo{year}{2020}\natexlab{}.
\newblock \showarticletitle{Avatar and {Sense} of {Embodiment}: {Studying} the {Relative} {Preference} {Between} {Appearance}, {Control} and {Point} of {View}}.
\newblock \bibinfo{journal}{\emph{IEEE Transactions on Visualization and Computer Graphics}} \bibinfo{volume}{26}, \bibinfo{number}{5} (\bibinfo{date}{May} \bibinfo{year}{2020}), \bibinfo{pages}{2062--2072}.
\newblock
\showISSN{1077-2626, 1941-0506, 2160-9306}
\urldef\tempurl%
\url{https://doi.org/10.1109/TVCG.2020.2973077}
\showDOI{\tempurl}


\bibitem[\protect\citeauthoryear{Fribourg, Ogawa, Hoyet, Argelaguet, Narumi, Hirose, and Lecuyer}{Fribourg et~al\mbox{.}}{2021}]%
        {fribourg_virtual_2021}
\bibfield{author}{\bibinfo{person}{Rebecca Fribourg}, \bibinfo{person}{Nami Ogawa}, \bibinfo{person}{Ludovic Hoyet}, \bibinfo{person}{Ferran Argelaguet}, \bibinfo{person}{Takuji Narumi}, \bibinfo{person}{Michitaka Hirose}, {and} \bibinfo{person}{Anatole Lecuyer}.} \bibinfo{year}{2021}\natexlab{}.
\newblock \showarticletitle{Virtual {Co}-{Embodiment}: {Evaluation} of the {Sense} of {Agency} {While} {Sharing} the {Control} of a {Virtual} {Body} {Among} {Two} {Individuals}}.
\newblock \bibinfo{journal}{\emph{IEEE Transactions on Visualization and Computer Graphics}} \bibinfo{volume}{27}, \bibinfo{number}{10} (\bibinfo{date}{Oct.} \bibinfo{year}{2021}), \bibinfo{pages}{4023--4038}.
\newblock
\showISSN{1077-2626, 1941-0506, 2160-9306}
\urldef\tempurl%
\url{https://doi.org/10.1109/TVCG.2020.2999197}
\showDOI{\tempurl}


\bibitem[\protect\citeauthoryear{Froese, Iizuka, and Ikegami}{Froese et~al\mbox{.}}{2014a}]%
        {froese2014embodied}
\bibfield{author}{\bibinfo{person}{Tom Froese}, \bibinfo{person}{Hiroyuki Iizuka}, {and} \bibinfo{person}{Takashi Ikegami}.} \bibinfo{year}{2014}\natexlab{a}.
\newblock \showarticletitle{Embodied social interaction constitutes social cognition in pairs of humans: a minimalist virtual reality experiment}.
\newblock \bibinfo{journal}{\emph{Scientific reports}} \bibinfo{volume}{4}, \bibinfo{number}{1} (\bibinfo{year}{2014}), \bibinfo{pages}{3672}.
\newblock


\bibitem[\protect\citeauthoryear{Froese, Iizuka, and Ikegami}{Froese et~al\mbox{.}}{2014b}]%
        {froese2014using}
\bibfield{author}{\bibinfo{person}{Tom Froese}, \bibinfo{person}{Hiroyuki Iizuka}, {and} \bibinfo{person}{Takashi Ikegami}.} \bibinfo{year}{2014}\natexlab{b}.
\newblock \showarticletitle{Using minimal human-computer interfaces for studying the interactive development of social awareness}.
\newblock \bibinfo{journal}{\emph{Frontiers in psychology}}  \bibinfo{volume}{5} (\bibinfo{year}{2014}), \bibinfo{pages}{1061}.
\newblock


\bibitem[\protect\citeauthoryear{Gonzalez-Franco, Pizarro, Cermeron, Li, Thorn, Hutabarat, Tiwari, and Bermell-Garcia}{Gonzalez-Franco et~al\mbox{.}}{2017}]%
        {gonzalez-franco_immersive_2017}
\bibfield{author}{\bibinfo{person}{Mar Gonzalez-Franco}, \bibinfo{person}{Rodrigo Pizarro}, \bibinfo{person}{Julio Cermeron}, \bibinfo{person}{Katie Li}, \bibinfo{person}{Jacob Thorn}, \bibinfo{person}{Windo Hutabarat}, \bibinfo{person}{Ashutosh Tiwari}, {and} \bibinfo{person}{Pablo Bermell-Garcia}.} \bibinfo{year}{2017}\natexlab{}.
\newblock \showarticletitle{Immersive {Mixed} {Reality} for {Manufacturing} {Training}}.
\newblock \bibinfo{journal}{\emph{Frontiers in Robotics and AI}}  \bibinfo{volume}{4} (\bibinfo{year}{2017}).
\newblock
\showISSN{2296-9144}
\urldef\tempurl%
\url{https://www.frontiersin.org/articles/10.3389/frobt.2017.00003}
\showURL{%
\tempurl}


\bibitem[\protect\citeauthoryear{Hagiwara, Sugimoto, Inami, and Kitazaki}{Hagiwara et~al\mbox{.}}{2019}]%
        {hagiwara_shared_2019}
\bibfield{author}{\bibinfo{person}{Takayoshi Hagiwara}, \bibinfo{person}{Maki Sugimoto}, \bibinfo{person}{Masahiko Inami}, {and} \bibinfo{person}{Michiteru Kitazaki}.} \bibinfo{year}{2019}\natexlab{}.
\newblock \showarticletitle{Shared {Body} by {Action} {Integration} of {Two} {Persons}: {Body} {Ownership}, {Sense} of {Agency} and {Task} {Performance}}. In \bibinfo{booktitle}{\emph{2019 {IEEE} {Conference} on {Virtual} {Reality} and {3D} {User} {Interfaces} ({VR})}}. \bibinfo{publisher}{IEEE}, \bibinfo{address}{Osaka, Japan}, \bibinfo{pages}{954--955}.
\newblock
\showISBNx{978-1-72811-377-7}
\urldef\tempurl%
\url{https://doi.org/10.1109/VR.2019.8798222}
\showDOI{\tempurl}


\bibitem[\protect\citeauthoryear{Hamad and Jia}{Hamad and Jia}{2022}]%
        {Hamad2022}
\bibfield{author}{\bibinfo{person}{Ayah Hamad} {and} \bibinfo{person}{Bochen Jia}.} \bibinfo{year}{2022}\natexlab{}.
\newblock \showarticletitle{How Virtual Reality Technology Has Changed Our Lives: An Overview of the Current and Potential Applications and Limitations}.
\newblock \bibinfo{journal}{\emph{International Journal of Environmental Research and Public Health}} \bibinfo{volume}{19}, \bibinfo{number}{18} (\bibinfo{year}{2022}).
\newblock
\showISSN{1660-4601}
\urldef\tempurl%
\url{https://doi.org/10.3390/ijerph191811278}
\showDOI{\tempurl}


\bibitem[\protect\citeauthoryear{Hapuarachchi, Hagiwara, Ganesh, and Kitazaki}{Hapuarachchi et~al\mbox{.}}{2023}]%
        {hapuarachchi_effect_2023}
\bibfield{author}{\bibinfo{person}{Harin Hapuarachchi}, \bibinfo{person}{Takayoshi Hagiwara}, \bibinfo{person}{Gowrishankar Ganesh}, {and} \bibinfo{person}{Michiteru Kitazaki}.} \bibinfo{year}{2023}\natexlab{}.
\newblock \showarticletitle{Effect of connection induced upper body movements on embodiment towards a limb controlled by another during virtual co-embodiment}.
\newblock \bibinfo{journal}{\emph{PLOS ONE}} \bibinfo{volume}{18}, \bibinfo{number}{1} (\bibinfo{date}{Jan.} \bibinfo{year}{2023}), \bibinfo{pages}{e0278022}.
\newblock
\showISSN{1932-6203}
\urldef\tempurl%
\url{https://doi.org/10.1371/journal.pone.0278022}
\showDOI{\tempurl}


\bibitem[\protect\citeauthoryear{Hapuarachchi and Kitazaki}{Hapuarachchi and Kitazaki}{2022}]%
        {hapuarachchi_knowing_2022}
\bibfield{author}{\bibinfo{person}{Harin Hapuarachchi} {and} \bibinfo{person}{Michiteru Kitazaki}.} \bibinfo{year}{2022}\natexlab{}.
\newblock \showarticletitle{Knowing the intention behind limb movements of a partner increases embodiment towards the limb of joint avatar}.
\newblock \bibinfo{journal}{\emph{Scientific Reports}} \bibinfo{volume}{12}, \bibinfo{number}{1} (\bibinfo{date}{July} \bibinfo{year}{2022}), \bibinfo{pages}{11453}.
\newblock
\showISSN{2045-2322}
\urldef\tempurl%
\url{https://doi.org/10.1038/s41598-022-15932-x}
\showDOI{\tempurl}


\bibitem[\protect\citeauthoryear{Harris, Arthur, Kearse, Olonilua, Hassan, De~Burgh, Wilson, and Vine}{Harris et~al\mbox{.}}{2023}]%
        {harris_exploring_2023}
\bibfield{author}{\bibinfo{person}{D.~J. Harris}, \bibinfo{person}{T. Arthur}, \bibinfo{person}{J. Kearse}, \bibinfo{person}{M. Olonilua}, \bibinfo{person}{E.~K. Hassan}, \bibinfo{person}{T.~C. De~Burgh}, \bibinfo{person}{M.~R. Wilson}, {and} \bibinfo{person}{S.~J. Vine}.} \bibinfo{year}{2023}\natexlab{}.
\newblock \showarticletitle{Exploring the role of virtual reality in military decision training}.
\newblock \bibinfo{journal}{\emph{Frontiers in Virtual Reality}}  \bibinfo{volume}{4} (\bibinfo{year}{2023}).
\newblock
\showISSN{2673-4192}
\urldef\tempurl%
\url{https://www.frontiersin.org/articles/10.3389/frvir.2023.1165030}
\showURL{%
\tempurl}


\bibitem[\protect\citeauthoryear{Heidicker, Langbehn, and Steinicke}{Heidicker et~al\mbox{.}}{2017}]%
        {heidicker_influence_2017}
\bibfield{author}{\bibinfo{person}{Paul Heidicker}, \bibinfo{person}{Eike Langbehn}, {and} \bibinfo{person}{Frank Steinicke}.} \bibinfo{year}{2017}\natexlab{}.
\newblock \showarticletitle{Influence of avatar appearance on presence in social {VR}}. In \bibinfo{booktitle}{\emph{3D {User} {Interfaces} (3DUI), 2017 {IEEE} {Symposium} on}}. \bibinfo{publisher}{IEEE}, \bibinfo{address}{Los Angeles, CA, USA}, \bibinfo{pages}{233--234}.
\newblock


\bibitem[\protect\citeauthoryear{Inami, Uriu, Kashino, Yoshida, Saito, Maekawa, and Kitazaki}{Inami et~al\mbox{.}}{2022}]%
        {Inami2022}
\bibfield{author}{\bibinfo{person}{Masahiko Inami}, \bibinfo{person}{Daisuke Uriu}, \bibinfo{person}{Zendai Kashino}, \bibinfo{person}{Shigeo Yoshida}, \bibinfo{person}{Hiroto Saito}, \bibinfo{person}{Azumi Maekawa}, {and} \bibinfo{person}{Michiteru Kitazaki}.} \bibinfo{year}{2022}\natexlab{}.
\newblock \showarticletitle{Cyborgs, Human Augmentation, Cybernetics, and JIZAI Body}. In \bibinfo{booktitle}{\emph{Augmented Humans 2022}} (Kashiwa, Chiba, Japan) \emph{(\bibinfo{series}{AHs 2022})}. \bibinfo{publisher}{Association for Computing Machinery}, \bibinfo{address}{New York, NY, USA}, \bibinfo{pages}{230?242}.
\newblock
\showISBNx{9781450396325}
\urldef\tempurl%
\url{https://doi.org/10.1145/3519391.3519401}
\showDOI{\tempurl}


\bibitem[\protect\citeauthoryear{Jeunet, Albert, Argelaguet, and Lecuyer}{Jeunet et~al\mbox{.}}{2018}]%
        {jeunet_you_2018}
\bibfield{author}{\bibinfo{person}{Camille Jeunet}, \bibinfo{person}{Louis Albert}, \bibinfo{person}{Ferran Argelaguet}, {and} \bibinfo{person}{Anatole Lecuyer}.} \bibinfo{year}{2018}\natexlab{}.
\newblock \showarticletitle{“{Do} {You} {Feel} in {Control}?”: {Towards} {Novel} {Approaches} to {Characterise}, {Manipulate} and {Measure} the {Sense} of {Agency} in {Virtual} {Environments}}.
\newblock \bibinfo{journal}{\emph{IEEE Transactions on Visualization and Computer Graphics}} \bibinfo{volume}{24}, \bibinfo{number}{4} (\bibinfo{date}{April} \bibinfo{year}{2018}), \bibinfo{pages}{1486--1495}.
\newblock
\showISSN{1077-2626}
\urldef\tempurl%
\url{https://doi.org/10.1109/TVCG.2018.2794598}
\showDOI{\tempurl}


\bibitem[\protect\citeauthoryear{Juan, Elexpuru, Dias, Santos, and Amorim}{Juan et~al\mbox{.}}{2023}]%
        {juan2023immersive}
\bibfield{author}{\bibinfo{person}{M-Carmen Juan}, \bibinfo{person}{Julen Elexpuru}, \bibinfo{person}{Paulo Dias}, \bibinfo{person}{Beatriz~Sousa Santos}, {and} \bibinfo{person}{Paula Amorim}.} \bibinfo{year}{2023}\natexlab{}.
\newblock \showarticletitle{Immersive virtual reality for upper limb rehabilitation: comparing hand and controller interaction}.
\newblock \bibinfo{journal}{\emph{Virtual Reality}} \bibinfo{volume}{27}, \bibinfo{number}{2} (\bibinfo{year}{2023}), \bibinfo{pages}{1157--1171}.
\newblock


\bibitem[\protect\citeauthoryear{Jung, Karki, Slutter, and Lindeman}{Jung et~al\mbox{.}}{2021}]%
        {jung_use_2021}
\bibfield{author}{\bibinfo{person}{Sungchul Jung}, \bibinfo{person}{Nawam Karki}, \bibinfo{person}{Max Slutter}, {and} \bibinfo{person}{Robert~W. Lindeman}.} \bibinfo{year}{2021}\natexlab{}.
\newblock \showarticletitle{On the {Use} of {Multi}-sensory {Cues} in {Symmetric} and {Asymmetric} {Shared} {Collaborative} {Virtual} {Spaces}}.
\newblock \bibinfo{journal}{\emph{Proceedings of the ACM on Human-Computer Interaction}} \bibinfo{volume}{5}, \bibinfo{number}{CSCW1} (\bibinfo{date}{April} \bibinfo{year}{2021}), \bibinfo{pages}{1--25}.
\newblock
\showISSN{2573-0142}
\urldef\tempurl%
\url{https://doi.org/10.1145/3449146}
\showDOI{\tempurl}


\bibitem[\protect\citeauthoryear{Keenaghan, Bowles, Crawfurd, Thurlbeck, Kentridge, and Cowie}{Keenaghan et~al\mbox{.}}{2020}]%
        {Keenaghan2020body}
\bibfield{author}{\bibinfo{person}{Samantha Keenaghan}, \bibinfo{person}{Lucy Bowles}, \bibinfo{person}{Georgina Crawfurd}, \bibinfo{person}{Simon Thurlbeck}, \bibinfo{person}{Robert~W. Kentridge}, {and} \bibinfo{person}{Dorothy Cowie}.} \bibinfo{year}{2020}\natexlab{}.
\newblock \showarticletitle{My body until proven otherwise: Exploring the time course of the full body illusion}.
\newblock \bibinfo{journal}{\emph{Consciousness and Cognition}}  \bibinfo{volume}{78} (\bibinfo{year}{2020}), \bibinfo{pages}{102882}.
\newblock
\showISSN{1053-8100}
\urldef\tempurl%
\url{https://doi.org/10.1016/j.concog.2020.102882}
\showDOI{\tempurl}


\bibitem[\protect\citeauthoryear{Kennedy, Lane, Berbaum, and Lilienthal}{Kennedy et~al\mbox{.}}{1993}]%
        {kennedy_simulator_1993}
\bibfield{author}{\bibinfo{person}{Robert~S. Kennedy}, \bibinfo{person}{Norman~E. Lane}, \bibinfo{person}{Kevin~S. Berbaum}, {and} \bibinfo{person}{Michael~G. Lilienthal}.} \bibinfo{year}{1993}\natexlab{}.
\newblock \showarticletitle{Simulator {Sickness} {Questionnaire}: {An} enhanced method for quantifying simulator sickness}.
\newblock \bibinfo{journal}{\emph{The International Journal of Aviation Psychology}} \bibinfo{volume}{3}, \bibinfo{number}{3} (\bibinfo{year}{1993}), \bibinfo{pages}{203--220}.
\newblock
\showISSN{1532-7108}
\urldef\tempurl%
\url{https://doi.org/10.1207/s15327108ijap0303_3}
\showDOI{\tempurl}
\newblock
\shownote{Place: US Publisher: Lawrence Erlbaum.}


\bibitem[\protect\citeauthoryear{Kilteni, Groten, and Slater}{Kilteni et~al\mbox{.}}{2012}]%
        {kilteni_sense_2012}
\bibfield{author}{\bibinfo{person}{Konstantina Kilteni}, \bibinfo{person}{Raphaela Groten}, {and} \bibinfo{person}{Mel Slater}.} \bibinfo{year}{2012}\natexlab{}.
\newblock \showarticletitle{The {Sense} of {Embodiment} in {Virtual} {Reality}}.
\newblock \bibinfo{journal}{\emph{Presence: Teleoperators and Virtual Environments}} \bibinfo{volume}{21}, \bibinfo{number}{4} (\bibinfo{date}{Nov.} \bibinfo{year}{2012}), \bibinfo{pages}{373--387}.
\newblock
\showISSN{1054-7460}
\urldef\tempurl%
\url{https://doi.org/10.1162/PRES_a_00124}
\showDOI{\tempurl}


\bibitem[\protect\citeauthoryear{Kodama, Mizuho, Hatada, Narumi, and Hirose}{Kodama et~al\mbox{.}}{2022}]%
        {kodama_enhancing_2022}
\bibfield{author}{\bibinfo{person}{Daiki Kodama}, \bibinfo{person}{Takato Mizuho}, \bibinfo{person}{Yuji Hatada}, \bibinfo{person}{Takuji Narumi}, {and} \bibinfo{person}{Michitaka Hirose}.} \bibinfo{year}{2022}\natexlab{}.
\newblock \showarticletitle{Enhancing the {Sense} of {Agency} by {Transitional} {Weight} {Control} in {Virtual} {Co}-{Embodiment}}. In \bibinfo{booktitle}{\emph{2022 {IEEE} {International} {Symposium} on {Mixed} and {Augmented} {Reality} ({ISMAR})}}. \bibinfo{publisher}{IEEE}, \bibinfo{address}{Singapore, Singapore}, \bibinfo{pages}{278--286}.
\newblock
\showISBNx{978-1-66545-325-7}
\urldef\tempurl%
\url{https://doi.org/10.1109/ISMAR55827.2022.00043}
\showDOI{\tempurl}


\bibitem[\protect\citeauthoryear{Kodama, Mizuho, Hatada, Narumi, and Hirose}{Kodama et~al\mbox{.}}{2023}]%
        {kodama_effects_2023}
\bibfield{author}{\bibinfo{person}{Daiki Kodama}, \bibinfo{person}{Takato Mizuho}, \bibinfo{person}{Yuji Hatada}, \bibinfo{person}{Takuji Narumi}, {and} \bibinfo{person}{Michitaka Hirose}.} \bibinfo{year}{2023}\natexlab{}.
\newblock \showarticletitle{Effects of {Collaborative} {Training} {Using} {Virtual} {Co}-embodiment on {Motor} {Skill} {Learning}}.
\newblock \bibinfo{journal}{\emph{IEEE Transactions on Visualization and Computer Graphics}} \bibinfo{volume}{29}, \bibinfo{number}{5} (\bibinfo{date}{May} \bibinfo{year}{2023}), \bibinfo{pages}{2304--2314}.
\newblock
\showISSN{1077-2626, 1941-0506, 2160-9306}
\urldef\tempurl%
\url{https://doi.org/10.1109/TVCG.2023.3247112}
\showDOI{\tempurl}


\bibitem[\protect\citeauthoryear{Kojima, Froese, Oka, Iizuka, and Ikegami}{Kojima et~al\mbox{.}}{2017}]%
        {kojima2017sensorimotor}
\bibfield{author}{\bibinfo{person}{Hiroki Kojima}, \bibinfo{person}{Tom Froese}, \bibinfo{person}{Mizuki Oka}, \bibinfo{person}{Hiroyuki Iizuka}, {and} \bibinfo{person}{Takashi Ikegami}.} \bibinfo{year}{2017}\natexlab{}.
\newblock \showarticletitle{A sensorimotor signature of the transition to conscious social perception: co-regulation of active and passive touch}.
\newblock \bibinfo{journal}{\emph{Frontiers in Psychology}}  \bibinfo{volume}{8} (\bibinfo{year}{2017}), \bibinfo{pages}{1778}.
\newblock


\bibitem[\protect\citeauthoryear{Lazar, Thompson, and Demiris}{Lazar et~al\mbox{.}}{2014}]%
        {amanda2014rt}
\bibfield{author}{\bibinfo{person}{Amanda Lazar}, \bibinfo{person}{Hilaire Thompson}, {and} \bibinfo{person}{George Demiris}.} \bibinfo{year}{2014}\natexlab{}.
\newblock \showarticletitle{A Systematic Review of the Use of Technology for Reminiscence Therapy}.
\newblock \bibinfo{journal}{\emph{Health Education \& Behavior}} \bibinfo{volume}{41}, \bibinfo{number}{1\_suppl} (\bibinfo{year}{2014}), \bibinfo{pages}{51S--61S}.
\newblock
\urldef\tempurl%
\url{https://doi.org/10.1177/1090198114537067}
\showDOI{\tempurl}
\showeprint{https://doi.org/10.1177/1090198114537067}
\newblock
\shownote{PMID: 25274711.}


\bibitem[\protect\citeauthoryear{Lenay}{Lenay}{2021}]%
        {lenay2021perceiving}
\bibfield{author}{\bibinfo{person}{Charles Lenay}.} \bibinfo{year}{2021}\natexlab{}.
\newblock \showarticletitle{Perceiving at a distance: enaction, exteriority and possibility--a tribute to John Stewart}.
\newblock \bibinfo{journal}{\emph{Adaptive Behavior}} \bibinfo{volume}{29}, \bibinfo{number}{5} (\bibinfo{year}{2021}), \bibinfo{pages}{485--503}.
\newblock


\bibitem[\protect\citeauthoryear{Lenay, Stewart, Rohde, and Amar}{Lenay et~al\mbox{.}}{2011}]%
        {lenay_you_2011}
\bibfield{author}{\bibinfo{person}{Charles Lenay}, \bibinfo{person}{John Stewart}, \bibinfo{person}{Marieke Rohde}, {and} \bibinfo{person}{Amal~Ali Amar}.} \bibinfo{year}{2011}\natexlab{}.
\newblock \showarticletitle{“{You} never fail to surprise me”: the hallmark of the {Other}: {Experimental} study and simulations of perceptual crossing}.
\newblock \bibinfo{journal}{\emph{Interaction Studies. Social Behaviour and Communication in Biological and Artificial Systems}} \bibinfo{volume}{12}, \bibinfo{number}{3} (\bibinfo{date}{Nov.} \bibinfo{year}{2011}), \bibinfo{pages}{373--396}.
\newblock
\showISSN{1572-0373, 1572-0381}
\urldef\tempurl%
\url{https://doi.org/10.1075/is.12.3.01len}
\showDOI{\tempurl}


\bibitem[\protect\citeauthoryear{Li, Kong, R\"{o}ggla, De~Simone, Ananthanarayan, de~Ridder, El~Ali, and Cesar}{Li et~al\mbox{.}}{2019}]%
        {Li2019a}
\bibfield{author}{\bibinfo{person}{Jie Li}, \bibinfo{person}{Yiping Kong}, \bibinfo{person}{Thomas R\"{o}ggla}, \bibinfo{person}{Francesca De~Simone}, \bibinfo{person}{Swamy Ananthanarayan}, \bibinfo{person}{Huib de Ridder}, \bibinfo{person}{Abdallah El~Ali}, {and} \bibinfo{person}{Pablo Cesar}.} \bibinfo{year}{2019}\natexlab{}.
\newblock \showarticletitle{Measuring and Understanding Photo Sharing Experiences in Social Virtual Reality}. In \bibinfo{booktitle}{\emph{Proceedings of the 2019 CHI Conference on Human Factors in Computing Systems}} (Glasgow, Scotland Uk) \emph{(\bibinfo{series}{CHI '19})}. \bibinfo{publisher}{ACM}, \bibinfo{address}{New York, NY, USA}, \bibinfo{pages}{1?14}.
\newblock
\showISBNx{9781450359702}
\urldef\tempurl%
\url{https://doi.org/10.1145/3290605.3300897}
\showDOI{\tempurl}


\bibitem[\protect\citeauthoryear{Lim and Reeves}{Lim and Reeves}{2009}]%
        {Lim2009}
\bibfield{author}{\bibinfo{person}{Sohye Lim} {and} \bibinfo{person}{Byron Reeves}.} \bibinfo{year}{2009}\natexlab{}.
\newblock \showarticletitle{Being in the Game: Effects of Avatar Choice and Point of View on Psychophysiological Responses During Play}.
\newblock \bibinfo{journal}{\emph{Media Psychology}} \bibinfo{volume}{12}, \bibinfo{number}{4} (\bibinfo{year}{2009}), \bibinfo{pages}{348--370}.
\newblock
\urldef\tempurl%
\url{https://doi.org/10.1080/15213260903287242}
\showDOI{\tempurl}
\showeprint{https://doi.org/10.1080/15213260903287242}


\bibitem[\protect\citeauthoryear{Lin, Kuo, Lin, Su, Lin, and Hsu}{Lin et~al\mbox{.}}{2021}]%
        {lin2021development}
\bibfield{author}{\bibinfo{person}{Che-Wei Lin}, \bibinfo{person}{Li-Chieh Kuo}, \bibinfo{person}{Yu-Ching Lin}, \bibinfo{person}{Fong-Chin Su}, \bibinfo{person}{Yu-An Lin}, {and} \bibinfo{person}{Hsiu-Yun Hsu}.} \bibinfo{year}{2021}\natexlab{}.
\newblock \showarticletitle{Development and testing of a virtual reality mirror therapy system for the sensorimotor performance of upper extremity: A pilot randomized controlled trial}.
\newblock \bibinfo{journal}{\emph{IEEE Access}}  \bibinfo{volume}{9} (\bibinfo{year}{2021}), \bibinfo{pages}{14725--14734}.
\newblock


\bibitem[\protect\citeauthoryear{Mekbib, Debeli, Zhang, Fang, Shao, Yang, Han, Jiang, Zhu, Zhao, et~al\mbox{.}}{Mekbib et~al\mbox{.}}{2021}]%
        {mekbib2021novel}
\bibfield{author}{\bibinfo{person}{Destaw~B Mekbib}, \bibinfo{person}{Dereje~Kebebew Debeli}, \bibinfo{person}{Li Zhang}, \bibinfo{person}{Shan Fang}, \bibinfo{person}{Yuling Shao}, \bibinfo{person}{Wei Yang}, \bibinfo{person}{Jiawei Han}, \bibinfo{person}{Hongjie Jiang}, \bibinfo{person}{Junming Zhu}, \bibinfo{person}{Zhiyong Zhao}, {et~al\mbox{.}}} \bibinfo{year}{2021}\natexlab{}.
\newblock \showarticletitle{A novel fully immersive virtual reality environment for upper extremity rehabilitation in patients with stroke}.
\newblock \bibinfo{journal}{\emph{Annals of the New York Academy of Sciences}} \bibinfo{volume}{1493}, \bibinfo{number}{1} (\bibinfo{year}{2021}), \bibinfo{pages}{75--89}.
\newblock


\bibitem[\protect\citeauthoryear{Melo, Gon{\c{c}}alves, Monteiro, Coelho, Vasconcelos-Raposo, and Bessa}{Melo et~al\mbox{.}}{2020}]%
        {melo2020multisensory}
\bibfield{author}{\bibinfo{person}{Miguel Melo}, \bibinfo{person}{Guilherme Gon{\c{c}}alves}, \bibinfo{person}{Pedro Monteiro}, \bibinfo{person}{Hugo Coelho}, \bibinfo{person}{Jos{\'e} Vasconcelos-Raposo}, {and} \bibinfo{person}{Maximino Bessa}.} \bibinfo{year}{2020}\natexlab{}.
\newblock \showarticletitle{Do multisensory stimuli benefit the virtual reality experience? A systematic review}.
\newblock \bibinfo{journal}{\emph{IEEE Transactions on Visualization and Computer Graphics}} \bibinfo{volume}{28}, \bibinfo{number}{2} (\bibinfo{year}{2020}), \bibinfo{pages}{1428--1442}.
\newblock


\bibitem[\protect\citeauthoryear{Moll and Salln{\"a}s}{Moll and Salln{\"a}s}{2009}]%
        {moll2009communicative}
\bibfield{author}{\bibinfo{person}{Jonas Moll} {and} \bibinfo{person}{Eva-Lotta Salln{\"a}s}.} \bibinfo{year}{2009}\natexlab{}.
\newblock \showarticletitle{Communicative functions of haptic feedback}. In \bibinfo{booktitle}{\emph{International Conference on Haptic and Audio Interaction Design}}. Springer, \bibinfo{publisher}{Springer}, \bibinfo{address}{Berlin, Heidelberg}, \bibinfo{pages}{1--10}.
\newblock


\bibitem[\protect\citeauthoryear{Mueller, Lopes, Strohmeier, Ju, Seim, Weigel, Nanayakkara, Obrist, Li, Delfa, Nishida, Gerber, Svanaes, Grudin, Greuter, Kunze, Erickson, Greenspan, Inami, Marshall, Reiterer, Wolf, Meyer, Schiphorst, Wang, and Maes}{Mueller et~al\mbox{.}}{2020}]%
        {Mueller2020}
\bibfield{author}{\bibinfo{person}{Florian~Floyd Mueller}, \bibinfo{person}{Pedro Lopes}, \bibinfo{person}{Paul Strohmeier}, \bibinfo{person}{Wendy Ju}, \bibinfo{person}{Caitlyn Seim}, \bibinfo{person}{Martin Weigel}, \bibinfo{person}{Suranga Nanayakkara}, \bibinfo{person}{Marianna Obrist}, \bibinfo{person}{Zhuying Li}, \bibinfo{person}{Joseph Delfa}, \bibinfo{person}{Jun Nishida}, \bibinfo{person}{Elizabeth~M. Gerber}, \bibinfo{person}{Dag Svanaes}, \bibinfo{person}{Jonathan Grudin}, \bibinfo{person}{Stefan Greuter}, \bibinfo{person}{Kai Kunze}, \bibinfo{person}{Thomas Erickson}, \bibinfo{person}{Steven Greenspan}, \bibinfo{person}{Masahiko Inami}, \bibinfo{person}{Joe Marshall}, \bibinfo{person}{Harald Reiterer}, \bibinfo{person}{Katrin Wolf}, \bibinfo{person}{Jochen Meyer}, \bibinfo{person}{Thecla Schiphorst}, \bibinfo{person}{Dakuo Wang}, {and} \bibinfo{person}{Pattie Maes}.} \bibinfo{year}{2020}\natexlab{}.
\newblock \showarticletitle{Next Steps for Human-Computer Integration}. In \bibinfo{booktitle}{\emph{Proceedings of the 2020 CHI Conference on Human Factors in Computing Systems}} (Honolulu, HI, USA) \emph{(\bibinfo{series}{CHI '20})}. \bibinfo{publisher}{Association for Computing Machinery}, \bibinfo{address}{New York, NY, USA}, \bibinfo{pages}{1?15}.
\newblock
\showISBNx{9781450367080}
\urldef\tempurl%
\url{https://doi.org/10.1145/3313831.3376242}
\showDOI{\tempurl}


\bibitem[\protect\citeauthoryear{Ogawa, Narumi, and Hirose}{Ogawa et~al\mbox{.}}{2019}]%
        {ogawa_virtual_2019}
\bibfield{author}{\bibinfo{person}{Nami Ogawa}, \bibinfo{person}{Takuji Narumi}, {and} \bibinfo{person}{Michitaka Hirose}.} \bibinfo{year}{2019}\natexlab{}.
\newblock \showarticletitle{Virtual {Hand} {Realism} {Affects} {Object} {Size} {Perception} in {Body}-{Based} {Scaling}}. In \bibinfo{booktitle}{\emph{2019 {IEEE} {Conference} on {Virtual} {Reality} and {3D} {User} {Interfaces} ({VR})}}. \bibinfo{publisher}{IEEE}, \bibinfo{address}{Osaka, Japan}, \bibinfo{pages}{519--528}.
\newblock
\showISBNx{978-1-72811-377-7}
\urldef\tempurl%
\url{https://doi.org/10.1109/VR.2019.8798040}
\showDOI{\tempurl}


\bibitem[\protect\citeauthoryear{Oh, Bailenson, and Welch}{Oh et~al\mbox{.}}{2018}]%
        {oh_systematic_2018}
\bibfield{author}{\bibinfo{person}{Catherine~S. Oh}, \bibinfo{person}{Jeremy~N. Bailenson}, {and} \bibinfo{person}{Gregory~F. Welch}.} \bibinfo{year}{2018}\natexlab{}.
\newblock \showarticletitle{A {Systematic} {Review} of {Social} {Presence}: {Definition}, {Antecedents}, and {Implications}}.
\newblock \bibinfo{journal}{\emph{Frontiers in Robotics and AI}}  \bibinfo{volume}{5} (\bibinfo{date}{Oct.} \bibinfo{year}{2018}), \bibinfo{pages}{114}.
\newblock
\showISSN{2296-9144}
\urldef\tempurl%
\url{https://doi.org/10.3389/frobt.2018.00114}
\showDOI{\tempurl}


\bibitem[\protect\citeauthoryear{Papacharissi}{Papacharissi}{2005}]%
        {papacharissi_real-virtual_2005}
\bibfield{author}{\bibinfo{person}{Zizi Papacharissi}.} \bibinfo{year}{2005}\natexlab{}.
\newblock \showarticletitle{The {Real}-{Virtual} {Dichotomy} in {Online} {Interaction}: {New} {Media} {Uses} and {Consequences} {Revisited}}.
\newblock \bibinfo{journal}{\emph{Annals of the International Communication Association}} \bibinfo{volume}{29}, \bibinfo{number}{1} (\bibinfo{date}{Jan.} \bibinfo{year}{2005}), \bibinfo{pages}{216--238}.
\newblock
\showISSN{2380-8985, 2380-8977}
\urldef\tempurl%
\url{https://doi.org/10.1080/23808985.2005.11679048}
\showDOI{\tempurl}


\bibitem[\protect\citeauthoryear{Peck and Gonzalez-Franco}{Peck and Gonzalez-Franco}{2021}]%
        {peck_avatar_2021}
\bibfield{author}{\bibinfo{person}{Tabitha~C. Peck} {and} \bibinfo{person}{Mar Gonzalez-Franco}.} \bibinfo{year}{2021}\natexlab{}.
\newblock \showarticletitle{Avatar {Embodiment}. {A} {Standardized} {Questionnaire}}.
\newblock \bibinfo{journal}{\emph{Frontiers in Virtual Reality}}  \bibinfo{volume}{1} (\bibinfo{date}{Feb.} \bibinfo{year}{2021}), \bibinfo{pages}{575943}.
\newblock
\showISSN{2673-4192}
\urldef\tempurl%
\url{https://doi.org/10.3389/frvir.2020.575943}
\showDOI{\tempurl}


\bibitem[\protect\citeauthoryear{Perry}{Perry}{2016}]%
        {Tekla2016}
\bibfield{author}{\bibinfo{person}{Tekla~S. Perry}.} \bibinfo{year}{2016}\natexlab{}.
\newblock \showarticletitle{Virtual reality goes social}.
\newblock \bibinfo{journal}{\emph{IEEE Spectrum}} \bibinfo{volume}{53}, \bibinfo{number}{1} (\bibinfo{year}{2016}), \bibinfo{pages}{56--57}.
\newblock
\urldef\tempurl%
\url{https://doi.org/10.1109/MSPEC.2016.7367470}
\showDOI{\tempurl}


\bibitem[\protect\citeauthoryear{Pimentel and Vinkers}{Pimentel and Vinkers}{2021}]%
        {pimentel_copresence_2021}
\bibfield{author}{\bibinfo{person}{Daniel Pimentel} {and} \bibinfo{person}{Charlotte Vinkers}.} \bibinfo{year}{2021}\natexlab{}.
\newblock \showarticletitle{Copresence {With} {Virtual} {Humans} in {Mixed} {Reality}: {The} {Impact} of {Contextual} {Responsiveness} on {Social} {Perceptions}}.
\newblock \bibinfo{journal}{\emph{Frontiers in Robotics and AI}}  \bibinfo{volume}{8} (\bibinfo{date}{April} \bibinfo{year}{2021}), \bibinfo{pages}{634520}.
\newblock
\showISSN{2296-9144}
\urldef\tempurl%
\url{https://doi.org/10.3389/frobt.2021.634520}
\showDOI{\tempurl}


\bibitem[\protect\citeauthoryear{Putze, Alexandrovsky, Putze, H\"{o}ffner, Smeddinck, and Malaka}{Putze et~al\mbox{.}}{2020}]%
        {Putze2020}
\bibfield{author}{\bibinfo{person}{Susanne Putze}, \bibinfo{person}{Dmitry Alexandrovsky}, \bibinfo{person}{Felix Putze}, \bibinfo{person}{Sebastian H\"{o}ffner}, \bibinfo{person}{Jan~David Smeddinck}, {and} \bibinfo{person}{Rainer Malaka}.} \bibinfo{year}{2020}\natexlab{}.
\newblock \showarticletitle{Breaking The Experience: Effects of Questionnaires in VR User Studies}. In \bibinfo{booktitle}{\emph{Proceedings of the 2020 CHI Conference on Human Factors in Computing Systems}} (Honolulu, HI, USA) \emph{(\bibinfo{series}{CHI '20})}. \bibinfo{publisher}{Association for Computing Machinery}, \bibinfo{address}{New York, NY, USA}, \bibinfo{pages}{1?15}.
\newblock
\showISBNx{9781450367080}
\urldef\tempurl%
\url{https://doi.org/10.1145/3313831.3376144}
\showDOI{\tempurl}


\bibitem[\protect\citeauthoryear{Rasch, Rusakov, Schmitz, and Müller}{Rasch et~al\mbox{.}}{2023}]%
        {rasch_going_2023}
\bibfield{author}{\bibinfo{person}{Julian Rasch}, \bibinfo{person}{Vladislav~Dmitrievic Rusakov}, \bibinfo{person}{Martin Schmitz}, {and} \bibinfo{person}{Florian Müller}.} \bibinfo{year}{2023}\natexlab{}.
\newblock \showarticletitle{Going, {Going}, {Gone}: {Exploring} {Intention} {Communication} for {Multi}-{User} {Locomotion} in {Virtual} {Reality}}. In \bibinfo{booktitle}{\emph{Proceedings of the 2023 {CHI} {Conference} on {Human} {Factors} in {Computing} {Systems}}}. \bibinfo{publisher}{ACM}, \bibinfo{address}{Hamburg Germany}, \bibinfo{pages}{1--13}.
\newblock
\showISBNx{978-1-4503-9421-5}
\urldef\tempurl%
\url{https://doi.org/10.1145/3544548.3581259}
\showDOI{\tempurl}


\bibitem[\protect\citeauthoryear{Rizvic, Young, Changa, Mijatovic, and Ivkovic-Kihic}{Rizvic et~al\mbox{.}}{2022}]%
        {rizvic_da_2022}
\bibfield{author}{\bibinfo{person}{Selma Rizvic}, \bibinfo{person}{Gregg Young}, \bibinfo{person}{Avinash Changa}, \bibinfo{person}{Bojan Mijatovic}, {and} \bibinfo{person}{Ivona Ivkovic-Kihic}.} \bibinfo{year}{2022}\natexlab{}.
\newblock \showarticletitle{{Da Vinci Effect - multiplayer Virtual Reality experience}}. In \bibinfo{booktitle}{\emph{Eurographics Workshop on Graphics and Cultural Heritage}}, \bibfield{editor}{\bibinfo{person}{Federico Ponchio} {and} \bibinfo{person}{Ruggero Pintus}} (Eds.). \bibinfo{publisher}{The Eurographics Association}, \bibinfo{address}{\mbox{}}.
\newblock
\showISBNx{978-3-03868-178-6}
\showISSN{2312-6124}
\urldef\tempurl%
\url{https://doi.org/10.2312/gch.20221229}
\showDOI{\tempurl}


\bibitem[\protect\citeauthoryear{Saarinen, Harjunen, Jasinskaja-Lahti, Jääskeläinen, and Ravaja}{Saarinen et~al\mbox{.}}{2021}]%
        {SAARINEN2021}
\bibfield{author}{\bibinfo{person}{Aino Saarinen}, \bibinfo{person}{Ville Harjunen}, \bibinfo{person}{Inga Jasinskaja-Lahti}, \bibinfo{person}{Iiro~P. Jääskeläinen}, {and} \bibinfo{person}{Niklas Ravaja}.} \bibinfo{year}{2021}\natexlab{}.
\newblock \showarticletitle{Social touch experience in different contexts: A review}.
\newblock \bibinfo{journal}{\emph{Neuroscience \& Biobehavioral Reviews}}  \bibinfo{volume}{131} (\bibinfo{year}{2021}), \bibinfo{pages}{360--372}.
\newblock
\showISSN{0149-7634}
\urldef\tempurl%
\url{https://doi.org/10.1016/j.neubiorev.2021.09.027}
\showDOI{\tempurl}


\bibitem[\protect\citeauthoryear{Schubert, Friedmann, and Regenbrecht}{Schubert et~al\mbox{.}}{2001}]%
        {schubert2001experience}
\bibfield{author}{\bibinfo{person}{Thomas Schubert}, \bibinfo{person}{Frank Friedmann}, {and} \bibinfo{person}{Holger Regenbrecht}.} \bibinfo{year}{2001}\natexlab{}.
\newblock \showarticletitle{The experience of presence: Factor analytic insights}.
\newblock \bibinfo{journal}{\emph{Presence: Teleoperators \& Virtual Environments}} \bibinfo{volume}{10}, \bibinfo{number}{3} (\bibinfo{year}{2001}), \bibinfo{pages}{266--281}.
\newblock


\bibitem[\protect\citeauthoryear{Schwind, Knierim, Tasci, Franczak, Haas, and Henze}{Schwind et~al\mbox{.}}{2017}]%
        {Schwind2017}
\bibfield{author}{\bibinfo{person}{Valentin Schwind}, \bibinfo{person}{Pascal Knierim}, \bibinfo{person}{Cagri Tasci}, \bibinfo{person}{Patrick Franczak}, \bibinfo{person}{Nico Haas}, {and} \bibinfo{person}{Niels Henze}.} \bibinfo{year}{2017}\natexlab{}.
\newblock \showarticletitle{"These Are Not My Hands!": Effect of Gender on the Perception of Avatar Hands in Virtual Reality}. In \bibinfo{booktitle}{\emph{Proceedings of the 2017 CHI Conference on Human Factors in Computing Systems}} (Denver, Colorado, USA) \emph{(\bibinfo{series}{CHI '17})}. \bibinfo{publisher}{Association for Computing Machinery}, \bibinfo{address}{New York, NY, USA}, \bibinfo{pages}{1577--1582}.
\newblock
\showISBNx{9781450346559}
\urldef\tempurl%
\url{https://doi.org/10.1145/3025453.3025602}
\showDOI{\tempurl}


\bibitem[\protect\citeauthoryear{Seifi, Zhang, and MacLean}{Seifi et~al\mbox{.}}{2015}]%
        {seifi2015vibviz}
\bibfield{author}{\bibinfo{person}{Hasti Seifi}, \bibinfo{person}{Kailun Zhang}, {and} \bibinfo{person}{Karon~E MacLean}.} \bibinfo{year}{2015}\natexlab{}.
\newblock \showarticletitle{VibViz: Organizing, visualizing and navigating vibration libraries}. In \bibinfo{booktitle}{\emph{2015 IEEE World Haptics Conference (WHC)}}. \bibinfo{publisher}{IEEE}, \bibinfo{address}{\mbox{}}, \bibinfo{pages}{254--259}.
\newblock


\bibitem[\protect\citeauthoryear{Smith and Neff}{Smith and Neff}{2018}]%
        {Harrison2018}
\bibfield{author}{\bibinfo{person}{Harrison~Jesse Smith} {and} \bibinfo{person}{Michael Neff}.} \bibinfo{year}{2018}\natexlab{}.
\newblock \showarticletitle{Communication Behavior in Embodied Virtual Reality}. In \bibinfo{booktitle}{\emph{Proceedings of the 2018 CHI Conference on Human Factors in Computing Systems}} (Montreal QC, Canada) \emph{(\bibinfo{series}{CHI '18})}. \bibinfo{publisher}{ACM}, \bibinfo{address}{New York, NY, USA}, \bibinfo{pages}{1--12}.
\newblock
\showISBNx{9781450356206}
\urldef\tempurl%
\url{https://doi.org/10.1145/3173574.3173863}
\showDOI{\tempurl}


\bibitem[\protect\citeauthoryear{Srinivasan and Basdogan}{Srinivasan and Basdogan}{1997}]%
        {srinivasan1997haptics}
\bibfield{author}{\bibinfo{person}{Mandayam~A Srinivasan} {and} \bibinfo{person}{Cagatay Basdogan}.} \bibinfo{year}{1997}\natexlab{}.
\newblock \showarticletitle{Haptics in virtual environments: Taxonomy, research status, and challenges}.
\newblock \bibinfo{journal}{\emph{Computers \& Graphics}} \bibinfo{volume}{21}, \bibinfo{number}{4} (\bibinfo{year}{1997}), \bibinfo{pages}{393--404}.
\newblock


\bibitem[\protect\citeauthoryear{Stephenson, Edwards, and Bayliss}{Stephenson et~al\mbox{.}}{2021}]%
        {Stephenson2021}
\bibfield{author}{\bibinfo{person}{Lisa~J. Stephenson}, \bibinfo{person}{S.~Gareth Edwards}, {and} \bibinfo{person}{Andrew~P. Bayliss}.} \bibinfo{year}{2021}\natexlab{}.
\newblock \showarticletitle{From Gaze Perception to Social Cognition: The Shared-Attention System}.
\newblock \bibinfo{journal}{\emph{Perspectives on Psychological Science}} \bibinfo{volume}{16}, \bibinfo{number}{3} (\bibinfo{year}{2021}), \bibinfo{pages}{553--576}.
\newblock
\urldef\tempurl%
\url{https://doi.org/10.1177/1745691620953773}
\showDOI{\tempurl}
\showeprint{https://doi.org/10.1177/1745691620953773}
\newblock
\shownote{PMID: 33567223.}


\bibitem[\protect\citeauthoryear{Sun, Shaikh, and Won}{Sun et~al\mbox{.}}{2019}]%
        {Sun2019}
\bibfield{author}{\bibinfo{person}{Yilu Sun}, \bibinfo{person}{Omar Shaikh}, {and} \bibinfo{person}{Andrea~Stevenson Won}.} \bibinfo{year}{2019}\natexlab{}.
\newblock \showarticletitle{Nonverbal synchrony in virtual reality}.
\newblock \bibinfo{journal}{\emph{PLOS ONE}} \bibinfo{volume}{14}, \bibinfo{number}{9} (\bibinfo{date}{09} \bibinfo{year}{2019}), \bibinfo{pages}{1--28}.
\newblock
\urldef\tempurl%
\url{https://doi.org/10.1371/journal.pone.0221803}
\showDOI{\tempurl}


\bibitem[\protect\citeauthoryear{Theodoropoulos, Stavropoulou, Papadopoulos, Platis, and Lepouras}{Theodoropoulos et~al\mbox{.}}{2023}]%
        {theodoropoulos_developing_2023}
\bibfield{author}{\bibinfo{person}{Anastasios Theodoropoulos}, \bibinfo{person}{Dimitra Stavropoulou}, \bibinfo{person}{Panagiotis Papadopoulos}, \bibinfo{person}{Nikos Platis}, {and} \bibinfo{person}{George Lepouras}.} \bibinfo{year}{2023}\natexlab{}.
\newblock \showarticletitle{Developing an {Interactive} {VR} {CAVE} for {Immersive} {Shared} {Gaming} {Experiences}}.
\newblock \bibinfo{journal}{\emph{Virtual Worlds}} \bibinfo{volume}{2}, \bibinfo{number}{2} (\bibinfo{date}{May} \bibinfo{year}{2023}), \bibinfo{pages}{162--181}.
\newblock
\showISSN{2813-2084}
\urldef\tempurl%
\url{https://doi.org/10.3390/virtualworlds2020010}
\showDOI{\tempurl}


\bibitem[\protect\citeauthoryear{Vesper, Abramova, B{\"u}tepage, Ciardo, Crossey, Effenberg, Hristova, Karlinsky, McEllin, Nijssen, et~al\mbox{.}}{Vesper et~al\mbox{.}}{2017}]%
        {vesper2017joint}
\bibfield{author}{\bibinfo{person}{Cordula Vesper}, \bibinfo{person}{Ekaterina Abramova}, \bibinfo{person}{Judith B{\"u}tepage}, \bibinfo{person}{Francesca Ciardo}, \bibinfo{person}{Benjamin Crossey}, \bibinfo{person}{Alfred Effenberg}, \bibinfo{person}{Dayana Hristova}, \bibinfo{person}{April Karlinsky}, \bibinfo{person}{Luke McEllin}, \bibinfo{person}{Sari~RR Nijssen}, {et~al\mbox{.}}} \bibinfo{year}{2017}\natexlab{}.
\newblock \showarticletitle{Joint action: Mental representations, shared information and general mechanisms for coordinating with others}.
\newblock \bibinfo{journal}{\emph{Frontiers in psychology}}  \bibinfo{volume}{7} (\bibinfo{year}{2017}), \bibinfo{pages}{2039}.
\newblock


\bibitem[\protect\citeauthoryear{Wee, Yap, and Lim}{Wee et~al\mbox{.}}{2021}]%
        {wee_haptic_2021}
\bibfield{author}{\bibinfo{person}{Chyanna Wee}, \bibinfo{person}{Kian~Meng Yap}, {and} \bibinfo{person}{Woan~Ning Lim}.} \bibinfo{year}{2021}\natexlab{}.
\newblock \showarticletitle{Haptic {Interfaces} for {Virtual} {Reality}: {Challenges} and {Research} {Directions}}.
\newblock \bibinfo{journal}{\emph{IEEE Access}}  \bibinfo{volume}{9} (\bibinfo{year}{2021}), \bibinfo{pages}{112145--112162}.
\newblock
\showISSN{2169-3536}
\urldef\tempurl%
\url{https://doi.org/10.1109/ACCESS.2021.3103598}
\showDOI{\tempurl}
\newblock
\shownote{Conference Name: IEEE Access.}


\bibitem[\protect\citeauthoryear{Wentzel, d'Eon, and Vogel}{Wentzel et~al\mbox{.}}{2020}]%
        {wentzel_improving_2020}
\bibfield{author}{\bibinfo{person}{Johann Wentzel}, \bibinfo{person}{Greg d'Eon}, {and} \bibinfo{person}{Daniel Vogel}.} \bibinfo{year}{2020}\natexlab{}.
\newblock \showarticletitle{Improving {Virtual} {Reality} {Ergonomics} {Through} {Reach}-{Bounded} {Non}-{Linear} {Input} {Amplification}}. In \bibinfo{booktitle}{\emph{Proceedings of the 2020 {CHI} {Conference} on {Human} {Factors} in {Computing} {Systems}}} \emph{(\bibinfo{series}{{CHI} '20})}. \bibinfo{publisher}{Association for Computing Machinery}, \bibinfo{address}{New York, NY, USA}, \bibinfo{pages}{1--12}.
\newblock
\showISBNx{978-1-4503-6708-0}
\urldef\tempurl%
\url{https://doi.org/10.1145/3313831.3376687}
\showDOI{\tempurl}


\bibitem[\protect\citeauthoryear{Wiener}{Wiener}{1948}]%
        {Wiener2008}
\bibfield{author}{\bibinfo{person}{Norbert Wiener}.} \bibinfo{year}{1948}\natexlab{}.
\newblock \bibinfo{booktitle}{\emph{Cybernetics: or Control and Communication in the Animal and the Machine} (\bibinfo{edition}{2} ed.)}.
\newblock \bibinfo{publisher}{MIT Press}, \bibinfo{address}{Cambridge, MA}.
\newblock


\bibitem[\protect\citeauthoryear{Wobbrock, Findlater, Gergle, and Higgins}{Wobbrock et~al\mbox{.}}{2011}]%
        {wobbrock_aligned_2011}
\bibfield{author}{\bibinfo{person}{Jacob~O. Wobbrock}, \bibinfo{person}{Leah Findlater}, \bibinfo{person}{Darren Gergle}, {and} \bibinfo{person}{James~J. Higgins}.} \bibinfo{year}{2011}\natexlab{}.
\newblock \showarticletitle{The aligned rank transform for nonparametric factorial analyses using only anova procedures}. In \bibinfo{booktitle}{\emph{Proceedings of the {SIGCHI} {Conference} on {Human} {Factors} in {Computing} {Systems}}} \emph{(\bibinfo{series}{{CHI} '11})}. \bibinfo{publisher}{Association for Computing Machinery}, \bibinfo{address}{New York, NY, USA}, \bibinfo{pages}{143--146}.
\newblock
\showISBNx{978-1-4503-0228-9}
\urldef\tempurl%
\url{https://doi.org/10.1145/1978942.1978963}
\showDOI{\tempurl}


\bibitem[\protect\citeauthoryear{Wohltjen, Toth, Boncz, and Wheatley}{Wohltjen et~al\mbox{.}}{2023}]%
        {wohltjen2023synchrony}
\bibfield{author}{\bibinfo{person}{Sophie Wohltjen}, \bibinfo{person}{Brigitta Toth}, \bibinfo{person}{Adam Boncz}, {and} \bibinfo{person}{Thalia Wheatley}.} \bibinfo{year}{2023}\natexlab{}.
\newblock \showarticletitle{Synchrony to a beat predicts synchrony with other minds}.
\newblock \bibinfo{journal}{\emph{Scientific Reports}} \bibinfo{volume}{13}, \bibinfo{number}{1} (\bibinfo{year}{2023}), \bibinfo{pages}{3591}.
\newblock


\bibitem[\protect\citeauthoryear{Yang, Kim, Jin, Gil, Koo, and Kim}{Yang et~al\mbox{.}}{2021}]%
        {Yang_2021}
\bibfield{author}{\bibinfo{person}{Tae-Heon Yang}, \bibinfo{person}{Jin~Ryong Kim}, \bibinfo{person}{Hanbit Jin}, \bibinfo{person}{Hyunjae Gil}, \bibinfo{person}{Jeong-Hoi Koo}, {and} \bibinfo{person}{Hye~Jin Kim}.} \bibinfo{year}{2021}\natexlab{}.
\newblock \showarticletitle{Recent Advances and Opportunities of Active Materials for Haptic Technologies in Virtual and Augmented Reality}.
\newblock \bibinfo{journal}{\emph{Advanced Functional Materials}} \bibinfo{volume}{31}, \bibinfo{number}{39} (\bibinfo{year}{2021}), \bibinfo{pages}{2008831}.
\newblock
\urldef\tempurl%
\url{https://doi.org/10.1002/adfm.202008831}
\showDOI{\tempurl}
\showeprint{https://onlinelibrary.wiley.com/doi/pdf/10.1002/adfm.202008831}


\bibitem[\protect\citeauthoryear{Yang and Kim}{Yang and Kim}{2002}]%
        {yang_2002}
\bibfield{author}{\bibinfo{person}{Ungyeon Yang} {and} \bibinfo{person}{Gerard~Jounghyun Kim}.} \bibinfo{year}{2002}\natexlab{}.
\newblock \showarticletitle{Implementation and Evaluation of "Just Follow Me": An Immersive, VR-Based, Motion-Training System}.
\newblock \bibinfo{journal}{\emph{Presence: Teleoper. Virtual Environ.}} \bibinfo{volume}{11}, \bibinfo{number}{3} (\bibinfo{date}{jun} \bibinfo{year}{2002}), \bibinfo{pages}{304–323}.
\newblock
\showISSN{1054-7460}
\urldef\tempurl%
\url{https://doi.org/10.1162/105474602317473240}
\showDOI{\tempurl}


\bibitem[\protect\citeauthoryear{Ye, Liu, Hu, and Xia}{Ye et~al\mbox{.}}{2022}]%
        {ye_neural3points_2022}
\bibfield{author}{\bibinfo{person}{Yongjing Ye}, \bibinfo{person}{Libin Liu}, \bibinfo{person}{Lei Hu}, {and} \bibinfo{person}{Shihong Xia}.} \bibinfo{year}{2022}\natexlab{}.
\newblock \bibinfo{title}{{Neural3Points}: {Learning} to {Generate} {Physically} {Realistic} {Full}-body {Motion} for {Virtual} {Reality} {Users}}.
\newblock
\newblock
\urldef\tempurl%
\url{https://doi.org/10.48550/arXiv.2209.05753}
\showDOI{\tempurl}
\newblock
\shownote{arXiv:2209.05753 [cs].}


\bibitem[\protect\citeauthoryear{Zhang, Li, Xu, Li, Yang, Tong, and Guo}{Zhang et~al\mbox{.}}{2023}]%
        {zhang_remotetouch_2023}
\bibfield{author}{\bibinfo{person}{Yizhong Zhang}, \bibinfo{person}{Zhiqi Li}, \bibinfo{person}{Sicheng Xu}, \bibinfo{person}{Chong Li}, \bibinfo{person}{Jiaolong Yang}, \bibinfo{person}{Xin Tong}, {and} \bibinfo{person}{Baining Guo}.} \bibinfo{year}{2023}\natexlab{}.
\newblock \bibinfo{title}{{RemoteTouch}: {Enhancing} {Immersive} {3D} {Video} {Communication} with {Hand} {Touch}}.
\newblock
\newblock
\urldef\tempurl%
\url{http://arxiv.org/abs/2302.14365}
\showURL{%
\tempurl}
\newblock
\shownote{arXiv:2302.14365 [cs].}


\bibitem[\protect\citeauthoryear{Zhu, Sui, Shen, Zhu, Ali, Guo, and Wang}{Zhu et~al\mbox{.}}{2021}]%
        {zhu2021dementia}
\bibfield{author}{\bibinfo{person}{Shizhe Zhu}, \bibinfo{person}{Youxin Sui}, \bibinfo{person}{Ying Shen}, \bibinfo{person}{Yi Zhu}, \bibinfo{person}{Nawab Ali}, \bibinfo{person}{Chuan Guo}, {and} \bibinfo{person}{Tong Wang}.} \bibinfo{year}{2021}\natexlab{}.
\newblock \showarticletitle{Effects of Virtual Reality Intervention on Cognition and Motor Function in Older Adults With Mild Cognitive Impairment or Dementia: A Systematic Review and Meta-Analysis}.
\newblock \bibinfo{journal}{\emph{Frontiers in Aging Neuroscience}}  \bibinfo{volume}{13} (\bibinfo{year}{2021}).
\newblock
\showISSN{1663-4365}
\urldef\tempurl%
\url{https://doi.org/10.3389/fnagi.2021.586999}
\showDOI{\tempurl}


\end{thebibliography}

\end{document}